\documentclass[aps,rmp,twocolumn,floatfix]{revtex4}

\usepackage{graphicx}
\usepackage{bm}
\usepackage{url}

\bibpunct{[}{]}{,}{a}{,}{,}

\begin{document}

\title{Econophysics, Statistical Mechanics Approach to}

\author{Victor M.~Yakovenko}

\affiliation{Department of Physics, University of
  Maryland, College Park, Maryland 20742-4111, USA}

\date[]{{\bf arXiv:0709.3662}, v.1 September 23, 2007, v.4 August 3,
2008}

\begin{abstract}
This is a review article for \textit{Encyclopedia of Complexity and
System Science}, to be published by Springer
\url{http://refworks.springer.com/complexity/}.  The terms highlighted
in {\bf bold} in Sec.\ \ref{Sec:Definition} refer to other articles in
this \textit{Encyclopedia}.  This paper reviews statistical models for
money, wealth, and income distributions developed in the econophysics
literature since late 1990s.\\

\textsf{\normalsize ``Money, it's a gas.'' Pink Floyd}
\end{abstract}

\maketitle

\tableofcontents

%%%%%%%%%%%%%%%%%%%%%%%%%%%%%%%%%%%%%%%%%%%%%%%%%%%%%%%%%%%%%%%
\section*{Glossary}
\label{Sec:Glossary}
%%%%%%%%%%%%%%%%%%%%%%%%%%%%%%%%%%%%%%%%%%%%%%%%%%%%%%%%%%%%%%%

{\bf Probability density} $P(x)$ is defined so that the probability of
finding a random variable $x$ in the interval from $x$ to $x+dx$ is
equal to $P(x)\,dx$.

{\bf Cumulative probability} $C(x)$ is defined as the integral
$C(x)=\int_x^\infty P(x)\,dx$.  It gives the probability that the
random variable exceeds a given value $x$.

{\bf The Boltzmann-Gibbs distribution} gives the probability of
finding a physical system in a state with the energy $\varepsilon$.
Its probability density is given by the exponential function
(\ref{Gibbs}).

{\bf The Gamma distribution} has the probability density  given
by a product of an exponential function and a power-law function, as
in Eq.\ (\ref{Gamma}).

{\bf The Pareto distribution} has the probability density $P(x)\propto
1/x^{1+\alpha}$ and the cumulative probability $C(x)\propto
1/x^\alpha$ given by a power law.  These expressions apply only for
high enough values of $x$ and do not apply for $x\to0$.

{\bf The Lorenz curve} was introduced by the American economist Max Lorenz
to describe income and wealth inequality.  It is defined in terms of
two coordinates $x(r)$ and $y(r)$ given by Eq.\ (\ref{xy}).  The
horizontal coordinate $x(r)$ is the fraction of the population with
income below $r$, and the vertical coordinate $y(r)$ is the fraction
of income this population accounts for.  As $r$ changes from 0 to
$\infty$, $x$ and $y$ change from 0 to 1, parametrically defining a
curve in the $(x,y)$-plane.

{\bf The Gini coefficient} $G$ was introduced by the Italian
statistician Corrado Gini as a measure of inequality in a society.  It
is defined as the area between the Lorenz curve and the straight
diagonal line, divided by the area of the triangle beneath the
diagonal line.  For perfect equality (everybody has the same income or
wealth) $G=0$, and for total inequality (one person has all income or
wealth, and the rest have nothing) $G=1$.

{\bf The Fokker-Planck equation} is the partial differential equation
(\ref{diffusion}) that describes evolution in time $t$ of the
probability density $P(r,t)$ of a random variable $r$ experiencing
small random changes $\Delta r$ during short time intervals $\Delta
t$.  It is also known in mathematical literature as the Kolmogorov
forward equation.  The diffusion equation is an example of the
Fokker-Planck equation.

%%%%%%%%%%%%%%%%%%%%%%%%%%%%%%%%%%%%%%%%%%%%%%%%%%%%%%%%%%%%%%%
\section{Definition of the subject}
\label{Sec:Definition}
%%%%%%%%%%%%%%%%%%%%%%%%%%%%%%%%%%%%%%%%%%%%%%%%%%%%%%%%%%%%%%%

{\bf Econophysics} is an interdisciplinary research field applying
methods of statistical physics to problems in economics and finance.
The term ``econophysics'' was first introduced by the prominent
theoretical physicist Eugene Stanley in 1995 at the conference
\textit{Dynamics of Complex Systems}, which was held in Calcutta (now
known as Kolkata) as a satellite meeting to the STATPHYS--19
conference in China \cite{Chakrabarti-history,Carbone-2007}.  The term
appeared in print for the first time in the paper by Stanley
\textit{et al.}\ \cite{Stanley-1996} in the proceedings of the
Calcutta conference.  The paper presented a manifesto of the new
field, arguing that ``behavior of large numbers of humans (as
measured, e.g., by economic indices) might conform to analogs of the
scaling laws that have proved useful in describing systems composed of
large numbers of inanimate objects'' \cite{Stanley-1996}.  Soon the
first econophysics conferences were organized: \textit{International
Workshop on Econophysics}, Budapest, 1997 and \textit{International
Workshop on Econophysics and Statistical Finance}, Palermo, 1998
\cite{Carbone-2007}, and the book \textit{An Introduction to
Econophysics} \cite{Stanley-book} was published.

The term ``econophysics'' was introduced by analogy with similar
terms, such as astrophysics, geophysics, and biophysics, which
describe applications of physics to different fields.  Particularly
important is the parallel with biophysics, which studies living
creatures, which still obey the laws of physics.  It should be
emphasized that econophysics does not literally apply the laws of
physics, such as Newton's laws or quantum mechanics, to humans, but
rather uses mathematical methods developed in statistical physics to
study statistical properties of complex economic systems consisting of
a large number of humans.  So, it may be considered as a branch of
applied theory of probabilities.  However, statistical physics is
distinctly different from mathematical statistics in its focus,
methods, and results.

Originating from physics as a quantitative science, econophysics
emphasizes quantitative analysis of large amounts of economic and
financial data, which became increasingly available with the massive
introduction of computers and the Internet.  Econophysics distances
itself from the verbose, narrative, and ideological style of political
economy and is closer to {\bf econometrics} in its focus.  Studying
mathematical models of a large number of interacting economic agents,
econophysics has much common ground with the {\bf agent-based modeling
and simulation}.  Correspondingly, it distances itself from the
representative-agent approach of traditional economics, which, by
definition, ignores statistical and heterogeneous aspects of the
economy.

Two major directions in econophysics are applications to finance and
economics.  Applications to finance are described in a separate
article, {\bf Econophysics of Financial Markets}, in this
encyclopedia.  Observational aspects are covered in the article {\bf
Econophysics, Observational}.  The present article, {\bf Econophysics,
Statistical Mechanics Approach to}, concentrates primarily on
statistical distributions of money, wealth, and income among
interacting economic agents.

Another direction related to econophysics has been advocated by the
theoretical physicist Serge Galam since early 1980 under the name of
{\bf sociophysics} \cite{Galam-history}, with the first appearance of
the term in print in Ref.\ \cite{Galam-1982}.  It echoes the term
``physique sociale'' proposed in the nineteenth century by Auguste
Comte, the founder of sociology.  Unlike econophysics, the term
``sociophysics'' did not catch on when first introduced, but it is
coming back with the popularity of econophysics and active promotion
by some physicists
\cite{Stauffer-history,Schweitzer-2003,Weidlich-2000}.  While the
principles of both fields have a lot in common, econophysics focuses
on the narrower subject of economic behavior of humans, where more
quantitative data is available, whereas sociophysics studies a broader
range of social issues.  The boundary between econophysics and
sociophysics is not sharp, and the two fields enjoy a good rapport
\cite{Econo-Socio-book}.  A more detailed description of historical
development in presented in Sec.\ \ref{Sec:History}.

%%%%%%%%%%%%%%%%%%%%%%%%%%%%%%%%%%%%%%%%%%%%%%%%%%%%%%%%%%%%%%%
\section{Historical Introduction}
\label{Sec:History}
%%%%%%%%%%%%%%%%%%%%%%%%%%%%%%%%%%%%%%%%%%%%%%%%%%%%%%%%%%%%%%%

Statistical mechanics was developed in the second half of the
nineteenth century by James Clerk Maxwell, Ludwig Boltzmann, and
Josiah Willard Gibbs.  These physicists believed in the existence of
atoms and developed mathematical methods for describing their
statistical properties, such as the probability distribution of
velocities of molecules in a gas (the Maxwell-Boltzmann distribution)
and the general probability distribution of states with different
energies (the Boltzmann--Gibbs distribution).  There are interesting
connections between the development of statistical physics and
statistics of social phenomena, which were recently brought up by the
science journalist Philip Ball \cite{Ball-2002,Ball-book}.

Collection and study of ``social numbers'', such as the rates of
death, birth, and marriage, has been growing progressively since the
seventeenth century \cite[Ch.~3]{Ball-book}.  The term ``statistics''
was introduced in the eighteenth century to denote these studies
dealing with the civil ``states'', and its practitioners were called
``statists''.  Popularization of social statistics in the nineteenth
century is particularly accredited to the Belgian astronomer Adolphe
Quetelet.  Before the 1850s, statistics was considered an empirical
arm of political economy, but then it started to transform into a
general method of quantitative analysis suitable for all disciplines.
It stimulated physicists to develop statistical mechanics in the
second half of the nineteenth century.

Rudolf Clausius started development of the kinetic theory of gases,
but it was James Clerk Maxwell who made a decisive step of deriving
the probability distribution of velocities of molecules in a gas.
Historical studies show \cite[Ch.~3]{Ball-book} that, in developing
statistical mechanics, Maxwell was strongly influenced and encouraged
by the widespread popularity of social statistics at the time.  This
approach was further developed by Ludwig Boltzmann, who was very
explicit about its origins \cite[p. 69]{Ball-book}:
\begin{quote}
  ``The molecules are like individuals, \ldots\ and the properties of
  gases only remain unaltered, because the number of these molecules,
  which on the average have a given state, is constant.''
\end{quote}
In his book \textit{Popul\"are Schrifen} from 1905
\cite{Boltzmann-1905}, Boltzmann praises Josiah Willard Gibbs for
systematic development of statistical mechanics.  Then, Boltzmann says
(cited from \cite{Austrian}):
\begin{quote}
  ``This opens a broad perspective, if we do not only think of
  mechanical objects.  Let's consider to apply this method to the
  statistics of living beings, society, sociology and so forth.''
\end{quote}
(The author is grateful to Michael E.~Fisher for bringing this quote
to his attention.)

It is worth noting that many now-famous economists were originally
educated in physics and engineering.  Vilfredo Pareto earned a degree
in mathematical sciences and a doctorate in engineering.  Working as a
civil engineer, he collected statistics demonstrating that
distributions of income and wealth in a society follow a power law
\cite{Pareto}.  He later became a professor of economics at Lausanne,
where he replaced L\'eon Walras, also an engineer by education.  The
influential American economist Irving Fisher was a student of Gibbs.
However, most of the mathematical apparatus transferred to economics
from physics was that of Newtonian mechanics and classical
thermodynamics \cite{Mirowski}.  It culminated in the neoclassical
concept of mechanistic equilibrium where the ``forces'' of supply and
demand balance each other.  The more general concept of statistical
equilibrium largely eluded mainstream economics.

With time, both physics and economics became more formal and rigid
in their specializations, and the social origin of statistical physics
was forgotten.  The situation is well summarized by Philip Ball
\cite[p.~69]{Ball-book}:
\begin{quote}
  ``Today physicists regard the application of statistical mechanics
  to social phenomena as a new and risky venture.  Few, it seems,
  recall how the process originated the other way around, in the days
  when physical science and social science were the twin siblings of a
  mechanistic philosophy and when it was not in the least disreputable
  to invoke the habits of people to explain the habits of inanimate
  particles.''
\end{quote}

Some physicists and economists attempted to connect the two
disciplines during the twentieth century.  The theoretical physicist
Ettore Majorana argued in favor of applying the laws of statistical
physics to social phenomena in a paper published after his mysterious
disappearance \cite{Majorana-1942}.  The statistical physicist Elliott
Montroll co-authored the book \textit{Introduction to Quantitative
Aspects of Social Phenomena} \cite{Montroll-book}.  Several economists
applied statistical physics to economic problems
\cite{Follmer-1974,Blume-1993,Foley-1994,Durlauf-1997}.  An early
attempt to bring together the leading theoretical physicists and
economists at the Santa Fe Institute was not entirely successful
\cite{SantaFe-1988}.  However, by the late 1990s, the attempts to
apply statistical physics to social phenomena finally coalesced into
the robust movements of econophysics and sociophysics, as described in
Sec.\ \ref{Sec:Definition}.

The current standing of econophysics within the physics and economics
communities is mixed.  Although an entry on econophysics has appeared
in the \textit{New Palgrave Dictionary of Economics}
\cite{Rosser-2008a}, it is fair to say that econophysics is not
accepted yet by mainstream economics.  Nevertheless, a number of
open-minded, nontraditional economists have joined this movement, and
the number is growing.  Under these circumstances, econophysicists
have most of their papers published in physics journals.  The journal
\textit{Physica A: Statistical Mechanics and its Applications} emerged
as the leader in econophysics publications and has even attracted
submissions from some \textit{bona fide} economists.  The mainstream
physics community is generally sympathetic to econophysics, although
it is not uncommon for econophysics papers to be rejected by
\textit{Physical Review Letters} on the grounds that ``it is not
physics''.  There are regular conference in econophysics, such as
\textit{Applications of Physics in Financial Analysis} (sponsored by
the European Physical Society), \textit{Nikkei Econophysics
Symposium}, and \textit{Econophysics Colloquium}.  Econophysics
sessions are included in the annual meetings of physical societies and
statistical physics conferences.  The overlap with economists is the
strongest in the field of agent-based simulation.  Not surprisingly,
the conference series WEHIA/ESHIA, which deals with heterogeneous
interacting agents, regularly includes sessions on econophysics.

%%%%%%%%%%%%%%%%%%%%%%%%%%%%%%%%%%%%%%%%%%%%%%%%%%%%%%%%%%%%%%%
\section{Statistical Mechanics of Money Distribution}
\label{Sec:money}
%%%%%%%%%%%%%%%%%%%%%%%%%%%%%%%%%%%%%%%%%%%%%%%%%%%%%%%%%%%%%%%

When modern econophysics started in the middle of 1990s, its
attention was primarily focused on analysis of financial markets.
However, three influential papers
\cite{Yakovenko-2000,Chakraborti-2000,Bouchaud-2000}, dealing with the
subject of money and wealth distributions, were published in year
2000.  They started a new direction that is closer to economics than
finance and created an expanding wave of follow-up publications.  We
start reviewing this subject with Ref.\ \cite{Yakovenko-2000}, whose
results are the most closely related to the traditional statistical
mechanics in physics.

%%%%%%%%%%%%%%%%%%%%%%%%%%%%%%%%%%%%%%%%%%%%%%%%%%%%%%%%%%%%%%%
\subsection{The Boltzmann-Gibbs distribution of energy}
\label{Sec:BGphysics}
%%%%%%%%%%%%%%%%%%%%%%%%%%%%%%%%%%%%%%%%%%%%%%%%%%%%%%%%%%%%%%%

The fundamental law of equilibrium statistical mechanics is the
Boltzmann-Gibbs distribution.  It states that the probability
$P(\varepsilon)$ of finding a physical system or sub-system in a state
with the energy $\varepsilon$ is given by the exponential function
\begin{equation}
  P(\varepsilon)=c\,e^{-\varepsilon/T},
\label{Gibbs}
\end{equation}
where $T$ is the temperature, and $c$ is a normalizing constant
\cite{Wannier}. Here we set the Boltzmann constant $k_B$ to unity by
choosing the energy units for measuring the physical temperature $T$.
Then, the expectation value of any physical variable $x$ can be
obtained as
\begin{equation}
  \langle x\rangle=\frac{\sum_k x_ke^{-\varepsilon_k/T}}
  {\sum_k e^{-\varepsilon_k/T}},
\label{expectation}
\end{equation}
where the sum is taken over all states of the system.  Temperature is
equal to the average energy per particle:
$T\sim\langle\varepsilon\rangle$, up to a numerical coefficient of the
order of 1.

Eq.\ (\ref{Gibbs}) can be derived in different ways \cite{Wannier}.
All derivations involve the two main ingredients: statistical
character of the system and conservation of energy $\varepsilon$.  One
of the shortest derivations can be summarized as follows.  Let us
divide the system into two (generally unequal) parts.  Then, the total
energy is the sum of the parts:
$\varepsilon=\varepsilon_1+\varepsilon_2$, whereas the probability is
the product of probabilities:
$P(\varepsilon)=P(\varepsilon_1)\,P(\varepsilon_2)$.  The only
solution of these two equations is the exponential function
(\ref{Gibbs}).

A more sophisticated derivation, proposed by Boltzmann himself, uses
the concept of entropy.  Let us consider $N$ particles with the total
energy $E$.  Let us divide the energy axis into small intervals (bins)
of width $\Delta\varepsilon$ and count the number of particles
$N_k$ having the energies from $\varepsilon_k$ to
$\varepsilon_k+\Delta\varepsilon$.  The ratio $N_k/N=P_k$ gives the
probability for a particle to have the energy $\varepsilon_k$.  Let us
now calculate the multiplicity $W$, which is the number of
permutations of the particles between different energy bins such that
the occupation numbers of the bins do not change.  This quantity is
given by the combinatorial formula in terms of the factorials
\begin{equation}
  W=\frac{N!}{N_1!\,N_2!\,N_3!\,\ldots}.
\label{multiplicity}
\end{equation}
The logarithm of multiplicity is called the entropy $S=\ln W$.  In the
limit of large numbers, the entropy per particle can be written in the
following form using the Stirling approximation for the factorials
\begin{equation}
  \frac{S}{N}=-\sum_k \frac{N_k}{N}\ln\left(\frac{N_k}{N}\right)
  =-\sum_k P_k\ln P_k.
\label{entropy}
\end{equation}
Now we would like to find what distribution of particles between
different energy states has the highest entropy, i.e.,\ the highest
multiplicity, provided that the total energy of the system,
$E=\sum_kN_k\varepsilon_k$, has a fixed value.  Solution of this
problem can be easily obtained using the method of Lagrange
multipliers \cite{Wannier}, and the answer gives the exponential
distribution (\ref{Gibbs}).

The same result can be derived from the {\bf ergodic theory}, which
says that the many-body system occupies all possible states of a given
total energy with equal probabilities.  Then it is straightforward to
show \cite{ergodic-a,ergodic-b} that the probability distribution of
the energy of an individual particle is given by Eq.\ (\ref{Gibbs}).

%%%%%%%%%%%%%%%%%%%%%%%%%%%%%%%%%%%%%%%%%%%%%%%%%%%%%%%%%%%%%%%
\subsection{Conservation of money}
\label{Sec:conservation}
%%%%%%%%%%%%%%%%%%%%%%%%%%%%%%%%%%%%%%%%%%%%%%%%%%%%%%%%%%%%%%%

The derivations outlined in Sec.\ \ref{Sec:BGphysics} are very general
and use only the statistical character of the system and the
conservation of energy.  So, one may expect that the exponential
Boltzmann-Gibbs distribution (\ref{Gibbs}) may apply to other
statistical systems with a conserved quantity.

The economy is a big statistical system with millions of participating
agents, so it is a promising target for applications of statistical
mechanics.  Is there a conserved quantity in the economy?  The paper
\cite{Yakovenko-2000} argued that such a conserved quantity is money
$m$.  Indeed, the ordinary economic agents can only receive money from
and give money to other agents.  They are not permitted to
``manufacture'' money, e.g.,\ to print dollar bills.  When one agent
$i$ pays money $\Delta m$ to another agent $j$ for some goods or
services, the money balances of the agents change as follows
\begin{eqnarray}
  && m_i\;\rightarrow\; m_i'=m_i-\Delta m,
\nonumber \\
  && m_j\;\rightarrow\; m_j'=m_j+\Delta m.
\label{transfer}
\end{eqnarray}
The total amount of money of the two agents before and after
transaction remains the same
\begin{equation}
  m_i+m_j=m_i'+m_j',
\label{conservation}
\end{equation}
i.e.,\ there is a local conservation law for money.  The rule
(\ref{transfer}) for the transfer of money is analogous to the
transfer of energy from one molecule to another in molecular
collisions in a gas, and Eq.\ (\ref{conservation}) is analogous to
conservation of energy in such collisions.

Addressing some misunderstandings developed in economic literature
\cite{Anglin-2005,Lux-2005,Ormerod-2006,Lux-2008}, we should emphasize
that, in the model of Ref.\ \cite{Yakovenko-2000}, the transfer of
money from one agent to another happens voluntarily, as a payment for
goods and services in a market economy.  However, the model only keeps
track of money flow, but does not keep track of what kind of goods and
service are delivered.  One reason for this is that many goods, e.g.,\
food and other supplies, and most services, e.g.,\ getting a haircut
or going to a movie, are not tangible and disappear after consumption.
Because they are not conserved, and also because they are measured in
different physical units, it is not very practical to keep track of
them.  In contrast, money is measured in the same unit (within a given
country with a single currency) and is conserved in transactions, so
it is straightforward to keep track of money flow.

Unlike, ordinary economic agents, a central bank or a central
government can inject money into the economy.  This process is
analogous to an influx of energy into a system from external sources,
e.g.,\ the Earth receives energy from the Sun.  Dealing with these
situations, physicists start with an idealization of a closed system
in thermal equilibrium and then generalize to an open system subject
to an energy flux.  As long as the rate of money influx from central
sources is slow compared with relaxation processes in the economy and
does not cause hyperinflation, the system is in quasi-stationary
statistical equilibrium with slowly changing parameters.  This
situation is analogous to heating a kettle on a gas stove slowly,
where the kettle has a well-defined, but slowly increasing temperature
at any moment of time.

Another potential problem with conservation of money is debt.  This
issue is discussed in more detail in Sec.\ \ref{Sec:debt}.  As a
starting point, Ref.\ \cite{Yakovenko-2000} first considered simple
models, where debt is not permitted.  This means that money balances
of agents cannot go below zero: $m_i\geq0$ for all $i$.  Transaction
(\ref{transfer}) takes place only when an agent has enough money to
pay the price: $m_i\geq\Delta m$, otherwise the transaction does not
take place.  If an agent spends all money, the balance drops to zero
$m_i=0$, so the agent cannot buy any goods from other agents.
However, this agent can still produce goods or services, sell them to
other agents, and receive money for that.  In real life, cash balance
dropping to zero is not at all unusual for people who live from
paycheck to paycheck.

The conservation law is the key feature for the successful functioning
of money.  If the agents were permitted to ``manufacture'' money, they
would be printing money and buying all goods for nothing, which would
be a disaster.  The physical medium of money is not essential, as long
as the conservation law is enforced.  Money may be in the form of
paper cash, but today it is more often represented by digits in
computerized bank accounts.  The conservation law is the fundamental
principle of accounting, whether in the single-entry or the
double-entry form.  More discussion of banks and debt is given in
Sec.\ \ref{Sec:debt}.

%%%%%%%%%%%%%%%%%%%%%%%%%%%%%%%%%%%%%%%%%%%%%%%%%%%%%%%%%%%%%%%
\subsection{The Boltzmann-Gibbs distribution of money}
\label{Sec:BGmoney}
%%%%%%%%%%%%%%%%%%%%%%%%%%%%%%%%%%%%%%%%%%%%%%%%%%%%%%%%%%%%%%%

Having recognized the principle of money conservation, Ref.\
\cite{Yakovenko-2000} argued that the stationary distribution of money
should be given by the exponential Boltzmann-Gibbs function analogous
to Eq.\ (\ref{Gibbs})
\begin{equation}
  P(m)=c\,e^{-m/T_m}.
\label{money}
\end{equation}
Here $c$ is a normalizing constant, and $T_m$ is the ``money
temperature'', which is equal to the average amount of money per
agent: $T=\langle m\rangle=M/N$, where $M$ is the total money, and $N$
is the number of agents.

\begin{figure}[b]
\includegraphics[width=0.9\linewidth]{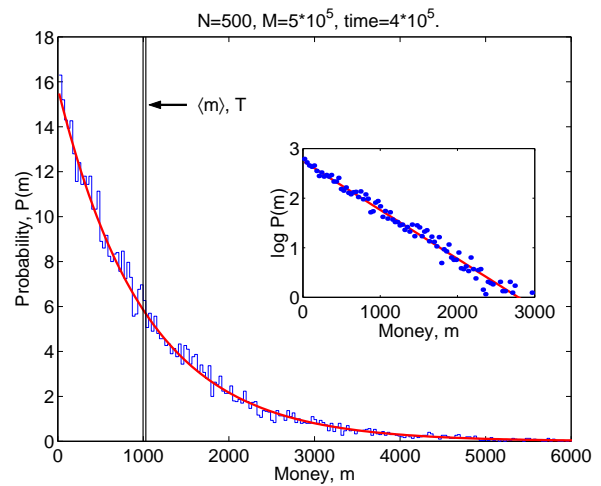}
\caption{\textit{Histogram and points:} Stationary probability
  distribution of money $P(m)$ obtained in agent-based computer
  simulations.  \textit{Solid curves:} Fits to the Boltzmann-Gibbs law
  (\ref{money}).  \textit{Vertical lines:} The initial distribution of
  money.  (Reproduced from Ref.\ \cite{Yakovenko-2000})}
\label{Fig:money}
\end{figure}

To verify this conjecture, Ref.\ \cite{Yakovenko-2000} performed
agent-based computer simulations of money transfers between agents.
Initially all agents were given the same amount of money, say, \$1000.
Then, a pair of agents $(i,j)$ was randomly selected, the amount
$\Delta m$ was transferred from one agent to another, and the process
was repeated many times.  Time evolution of the probability
distribution of money $P(m)$ can be seen in computer animation videos
at the Web pages \cite{animation,Wright-2007}.  After a transitory
period, money distribution converges to the stationary form shown in
Fig.\ \ref{Fig:money}.  As expected, the distribution is very well
fitted by the exponential function (\ref{money}).

%%%%%%%%%%%%%%%%%%%%%%%%%%%%%%%%%%%%%%%%%%%%%%%%%%%%%%%%%%%%%%%
%%%%%%%%%%%%%%%%%%%%%%%%%%%%%%%%%%%%%%%%%%%%%%%%%%%%%%%%%%%%%%%

Several different rules for $\Delta m$ were considered in Ref.\
\cite{Yakovenko-2000}.  In one model, the transferred amount was fixed
to a constant $\Delta m=\$1$.  Economically, it means that all agents
were selling their products for the same price $\Delta m=\$1$.
Computer animation \cite{animation} shows that the initial
distribution of money first broadens to a symmetric, Gaussian curve,
characteristic for a diffusion process.  Then, the distribution starts
to pile up around the $m=0$ state, which acts as the impenetrable
boundary, because of the imposed condition $m\geq0$.  As a result,
$P(m)$ becomes skewed (asymmetric) and eventually reaches the
stationary exponential shape, as shown in Fig.\ \ref{Fig:money}.  The
boundary at $m=0$ is analogous to the ground state energy in
statistical physics.  Without this boundary condition, the probability
distribution of money would not reach a stationary state.  Computer
animation \cite{animation,Wright-2007} also shows how the entropy of
money distribution, defined as $S/N=-\sum_k P(m_k)\ln P(m_k)$, grows
from the initial value $S=0$, when all agents have the same money, to
the maximal value at the statistical equilibrium.

While the model with $\Delta m=1$ is very simple and instructive, it
is not very realistic, because all prices are taken to be the same.
In another model considered in Ref.\ \cite{Yakovenko-2000}, $\Delta m$
in each transaction is taken to be a random fraction of the average
amount of money per agent, i.e.,\ $\Delta m=\nu(M/N)$, where $\nu$ is a
uniformly distributed random number between 0 and 1.  The random
distribution of $\Delta m$ is supposed to represent the wide variety
of prices for different products in the real economy.  It reflects the
fact that agents buy and consume many different types of products,
some of them simple and cheap, some sophisticated and expensive.
Moreover, different agents like to consume these products in different
quantities, so there is variation of paid amounts $\Delta m$, even
though the unit price of the same product is constant.  Computer
simulation of this model produces exactly the same stationary
distribution (\ref{money}), as in the first model.  Computer animation
for this model is also available on the Web page \cite{animation}.

The final distribution is universal despite different rules for
$\Delta m$.  To amplify this point further, Ref.\
\cite{Yakovenko-2000} also considered a toy model, where $\Delta m$
was taken to be a random fraction of the average amount of money of
the two agents: $\Delta m=\nu(m_i+m_j)/2$.  This model produced the
same stationary distribution (\ref{money}) as the two other models.

The pairwise models of money transfer are attractive in their
simplicity, but they represent a rather primitive market.  Modern
economy is dominated by big firms, which consist of many agents, so
Ref.\ \cite{Yakovenko-2000} also studied a model with firms.  One
agent at a time is appointed to become a ``firm''.  The firm borrows
capital $K$ from another agent and returns it with interest $hK$,
hires $L$ agents and pays them wages $W$, manufactures $Q$ items of a
product, sells them to $Q$ agents at price $R$, and receives profit
$F=RQ-LW-hK$.  All of these agents are randomly selected.  The
parameters of the model are optimized following a procedure from
economics textbooks \cite{McConnell}.  The aggregate demand-supply
curve for the product is taken in the form $R(Q)=v/Q^\eta$, where $Q$
is the quantity consumers would buy at price $R$, and $\eta$ and $v$
are some parameters.  The production function of the firm has the
traditional Cobb-Douglas form: $Q(L,K)=L^\chi K^{1-\chi}$, where
$\chi$ is a parameter.  Then the profit of the firm $F$ is maximized
with respect to $K$ and $L$.  The net result of the firm activity is a
many-body transfer of money, which still satisfies the conservation
law.  Computer simulation of this model generates the same exponential
distribution (\ref{money}), independently of the model parameters.
The reasons for the universality of the Boltzmann-Gibbs distribution
and its limitations are discussed from a different perspective in
Sec.\ \ref{Sec:+x}.

Well after the paper \cite{Yakovenko-2000} appeared, Italian
econophysicists \cite{Germano-2005} found that similar ideas had been
published earlier in obscure journals in Italian by Eleonora Bennati
\cite{Bennati-1988,Bennati-1993}.  They proposed calling these models
the Bennati-Dragulescu-Yakovenko (BDY) game \cite{Scalas-2006}.  The
Boltzmann distribution was independently applied to social sciences by
J\"urgen Mimkes using the Lagrange principle of maximization with
constraints \cite{Mimkes-2000,Mimkes-2005a}.  The exponential
distribution of money was also found in Ref.\ \cite{Shubik-1999} using
a Markov chain approach to strategic market games.  A long time ago,
Benoit Mandelbrot observed \cite[p 83]{Mandelbrot-1960}:
\begin{quote}
  ``There is a great temptation to consider the exchanges of money
  which occur in economic interaction as analogous to the exchanges of
  energy which occur in physical shocks between gas molecules.''
\end{quote}
He realized that this process should result in the exponential
distribution, by analogy with the barometric distribution of density
in the atmosphere.  However, he discarded this idea, because it does
not produce the Pareto power law, and proceeded to study the stable
L\'evy distributions.  Ironically, the actual economic data, discussed
in Secs.\ \ref{Sec:w-empirical} and \ref{Sec:r-data}, do show the
exponential distribution for the majority of the population.
Moreover, the data have finite variance, so the stable L\'evy
distributions are not applicable because of their infinite variance.

%%%%%%%%%%%%%%%%%%%%%%%%%%%%%%%%%%%%%%%%%%%%%%%%%%%%%%%%%%%%%%%
\subsection{Models with debt}
\label{Sec:debt}
%%%%%%%%%%%%%%%%%%%%%%%%%%%%%%%%%%%%%%%%%%%%%%%%%%%%%%%%%%%%%%%

\begin{figure}[b]
\includegraphics[width=0.9\linewidth]{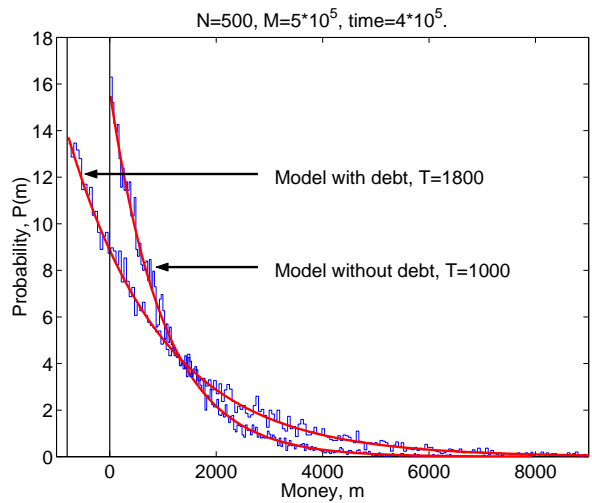}
\caption{\textit{Histograms:} Stationary distributions of money with
  and without debt.  The debt is limited to $m_d=800$. \textit{Solid
  curves:} Fits to the Boltzmann-Gibbs laws with the ``temperatures''
  $T=1800$ and $T=1000$. (Reproduced from Ref.\
  \cite{Yakovenko-2000})}
\label{Fig:debt}
\end{figure}

Now let us discuss how the results change when debt is permitted.
Debt may be considered as negative money.  When an agent borrows money
from a bank (considered here as a big reservoir of money), the cash
balance of the agent (positive money) increases, but the agent also
acquires a debt obligation (negative money), so the total balance (net
worth) of the agent remains the same, and the conservation law of
total money (positive and negative) is satisfied.  After spending some
cash, the agent still has the debt obligation, so the money balance of
the agent becomes negative.  Any stable economic system must have a
mechanism preventing unlimited borrowing and unlimited debt.
Otherwise, agents can buy any products without producing anything in
exchange by simply going into unlimited debt.  The exact mechanisms of
limiting debt in the real economy are complicated and obscured.  Ref.\
\cite{Yakovenko-2000} considered a simple model where the maximal debt
of any agent is limited by a certain amount $m_d$.  This means that
the boundary condition $m_i\geq0$ is now replaced by the condition
$m_i\geq-m_d$ for all agents $i$.  Setting interest rates on borrowed
money to be zero for simplicity, Ref.\ \cite{Yakovenko-2000} performed
computer simulations of the models described in Sec.\
\ref{Sec:BGmoney} with the new boundary condition.  The results are
shown in Fig.\ \ref{Fig:debt}.  Not surprisingly, the stationary money
distribution again has the exponential shape, but now with the new
boundary condition at $m=-m_d$ and the higher money temperature
$T_d=m_d+M/N$.  By allowing agents to go into debt up to $m_d$, we
effectively increase the amount of money available to each agent by
$m_d$.  So, the money temperature, which is equal to the average
amount of effectively available money per agent, increases.  A model
with non-zero interest rates was also studied in Ref.\
\cite{Yakovenko-2000}.

%%%%%%%%%%%%%%%%%%%%%%%%%%%%%%%%%%%%%%%%%%%%%%%%%%%%%%%%%%%%%%%
%%%%%%%%%%%%%%%%%%%%%%%%%%%%%%%%%%%%%%%%%%%%%%%%%%%%%%%%%%%%%%%

We see that debt does not violate the conservation law of money, but
rather modifies boundary conditions for $P(m)$.  When economics
textbooks describe how ``banks create money'' or ``debt creates
money'' \cite{McConnell}, they count only positive money (cash) as
money, but do not count liabilities (debt obligations) as negative
money.  With such a definition, money is not conserved.  However, if
we include debt obligations in the definition of money, then the
conservation law is restored.  This approach is in agreement with the
principles of double-entry accounting, which records both assets and
debts.  Debt obligations are as real as positive cash, as many
borrowers painfully realized in their experience.  A more detailed
study of positive and negative money and book-keeping from the point
of view of econophysics is presented in a series of papers by the
physicist Dieter Braun \cite{Braun-2001,Braun-2003a,Braun-2003b}.

\begin{figure}[b]
\includegraphics[width=\linewidth]{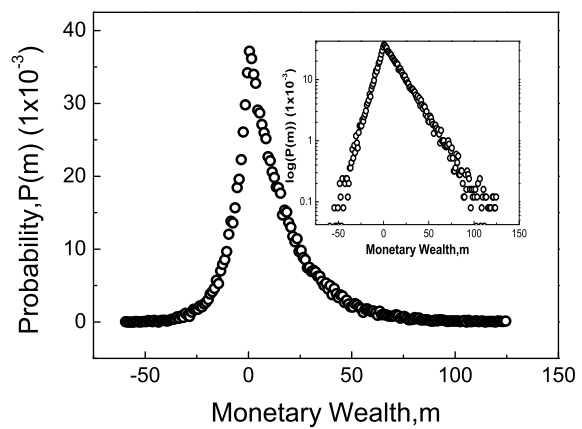}
\caption{The stationary distribution of money for the required reserve
  ratio $r=0.8$.  The distribution is exponential for positive and
  negative money with different ``temperatures'' $T_+$ and $T_-$, as
  illustrated by the inset on log-linear scale.  (Reproduced from
  Ref.\ \cite{ReserveRatio})}
\label{Fig:reserve}
\end{figure}

One way of limiting the total debt in the economy is the so-called
required reserve ratio $r$ \cite{McConnell}.  Every bank is required
by law to set aside a fraction $r$ of the money deposited into the
bank, and this reserved money cannot be loaned further.  If the
initial amount of money in the system (the money base) is $M_0$, then
with loans and borrowing the total amount of positive money available
to the agents increases to $M=M_0/r$, where the factor $1/r$ is called
the money multiplier \cite{McConnell}.  This is how ``banks create
money''.  Where does this extra money come from?  It comes from the
increase of the total debt in the system.  The maximal total debt is
equal to $D=M_0/r-M_0$ and is limited by the factor $r$.  When the
debt is maximal, the total amounts of positive, $M_0/r$, and negative,
$M_0(1-r)/r$, money circulate between the agents in the system, so
there are effectively two conservation laws for each of them
\cite{ReserveRatio}.  Thus, we expect to see the exponential
distributions of positive and negative money characterized by two
different temperatures: $T_+=M_0/rN$ and $T_-=M_0(1-r)/rN$.  This is
exactly what was found in computer simulations in Ref.\
\cite{ReserveRatio}, shown in Fig.\ \ref{Fig:reserve}.  Similar
two-sided distributions were also found in Ref.\ \cite{Braun-2003a}.

%%%%%%%%%%%%%%%%%%%%%%%%%%%%%%%%%%%%%%%%%%%%%%%%%%%%%%%%%%%%%%%
%%%%%%%%%%%%%%%%%%%%%%%%%%%%%%%%%%%%%%%%%%%%%%%%%%%%%%%%%%%%%%%

%%%%%%%%%%%%%%%%%%%%%%%%%%%%%%%%%%%%%%%%%%%%%%%%%%%%%%%%%%%%%%%
\subsection{Proportional money transfers and saving propensity}
\label{Sec:saving}
%%%%%%%%%%%%%%%%%%%%%%%%%%%%%%%%%%%%%%%%%%%%%%%%%%%%%%%%%%%%%%%

\begin{figure}[b]
\includegraphics[width=0.9\linewidth]{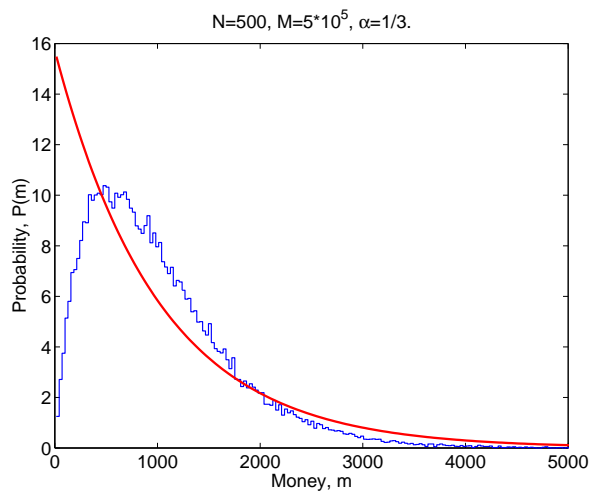}
\caption{\textit{Histogram:} Stationary probability distribution of
money in the multiplicative random exchange model (\ref{proportional})
for $\gamma=1/3$.  \textit{Solid curve:} The exponential
Boltzmann-Gibbs law.  (Reproduced from Ref.\ \cite{Yakovenko-2000})}
\label{Fig:Redner}
\end{figure}

In the models of money transfer considered thus far, the transferred
amount $\Delta m$ is typically independent of the money balances of
agents.  A different model was introduced in physics literature
earlier \cite{Redner-1998} under the name multiplicative asset
exchange model.  This model also satisfies the conservation law, but
the transferred amount of money is a fixed fraction $\gamma$ of the
payer's money in Eq.\ (\ref{transfer}):
\begin{equation}
  \Delta m=\gamma m_i.
\label{proportional}
\end{equation}
The stationary distribution of money in this model, shown in Fig.\
\ref{Fig:Redner} with an exponential function, is close, but not
exactly equal, to the Gamma distribution:
\begin{equation}
  P(m)=c\,m^\beta\,e^{-m/T}.
\label{Gamma}
\end{equation}
Eq.\ (\ref{Gamma}) differs from Eq.\ (\ref{money}) by the power-law
prefactor $m^\beta$.  From the Boltzmann kinetic equation (discussed
in more detail in Sec.\ \ref{Sec:+x}), Ref.\ \cite{Redner-1998}
derived a formula relating the parameters $\gamma$ and $\beta$ in
Eqs.\ (\ref{proportional}) and (\ref{Gamma}):
\begin{equation}
  \beta=-1-\ln2/\ln(1-\gamma).
\label{beta}
\end{equation}
When payers spend a relatively small fraction of their money
$\gamma<1/2$, Eq.\ (\ref{beta}) gives $\beta>0$, so the low-money
population is reduced and $P(m\to0)=0$, as shown in Fig.\
\ref{Fig:Redner}.

%%%%%%%%%%%%%%%%%%%%%%%%%%%%%%%%%%%%%%%%%%%%%%%%%%%%%%%%%%%%%%%
%%%%%%%%%%%%%%%%%%%%%%%%%%%%%%%%%%%%%%%%%%%%%%%%%%%%%%%%%%%%%%%

Later, the economist Thomas Lux brought to the attention of physicists
\cite{Lux-2005} that essentially the same model, called the inequality
process, had been introduced and studied much earlier by the
sociologist John Angle
\cite{Angle-1986,Angle-1992,Angle-1993,Angle-1996,Angle-2002}, see
also the review \cite{Angle-2006} for additional references.  While
Ref.\ \cite{Redner-1998} did not give much justification for the
proportionality law (\ref{proportional}), Angle \cite{Angle-1986}
connected this rule with the surplus theory of social stratification
\cite{Engels}, which argues that inequality in human society develops
when people can produce more than necessary for minimal subsistence.
This additional wealth (surplus) can be transferred from original
producers to other people, thus generating inequality.  In the first
paper by Angle \cite{Angle-1986}, the parameter $\gamma$ was randomly
distributed, and another parameter $\delta$ gave a higher probability
of winning to the agent with a higher money balance in Eq.\
(\ref{transfer}).  However, in the following papers, he simplified the
model to a fixed $\gamma$ (denoted as $\omega$ by Angle) and equal
probabilities of winning for higher- and lower-balance agents, which
makes it completely equivalent to the model of Ref.\
\cite{Redner-1998}.  Angle also considered a model
\cite{Angle-2002,Angle-2006} where groups of agents have different
values of $\gamma$, simulating the effect of education and other
``human capital''.  All of these models generate a Gamma-like
distribution, well approximated by Eq.\ (\ref{Gamma}).

Another model with an element of proportionality was proposed in Ref.\
\cite{Chakraborti-2000}.  (This paper originally appeared as a
follow-up preprint cond-mat/0004256 to the preprint cond-mat/0001432
of Ref.\ \cite{Yakovenko-2000}.)  In this model, the agents set aside
(save) some fraction of their money $\lambda m_i$, whereas the rest of
their money balance $(1-\lambda)m_i$ becomes available for random
exchanges.  Thus, the rule of exchange (\ref{transfer}) becomes
\begin{eqnarray}
  && m_i'=\lambda m_i + \xi(1-\lambda)(m_i+m_j),
\nonumber \\
  && m_j'=\lambda m_j + (1-\xi)(1-\lambda)(m_i+m_j).
\label{saving}
\end{eqnarray}
Here the coefficient $\lambda$ is called the saving propensity, and
the random variable $\xi$ is uniformly distributed between 0 and 1.
It was pointed out in Ref.\ \cite{Angle-2006} that, by the change of
notation $\lambda\to(1-\gamma)$, Eq.\ (\ref{saving}) can be
transformed to the same form as Eq.\ (\ref{proportional}), if the
random variable $\xi$ takes only discrete values 0 and 1.  Computer
simulations \cite{Chakraborti-2000} of the model (\ref{saving}) found
a stationary distribution close to the Gamma distribution
(\ref{Gamma}).  It was shown that the parameter $\beta$ is related to
the saving propensity $\lambda$ by the formula
$\beta=3\lambda/(1-\lambda)$
\cite{Patriarca-2004a,Patriarca-2004b,Germano-2005,Richmond-2005a}.
For $\lambda\neq0$, agents always keep some money, so their balances
never go to zero and $P(m\to0)=0$, whereas for $\lambda=0$ the
distribution becomes exponential.

In the subsequent papers by the Kolkata school
\cite{Chakrabarti-history} and related papers, the case of random
saving propensity was studied.  In these models, the agents are
assigned random parameters $\lambda$ drawn from a uniform distribution
between 0 and 1 \cite{Chakrabarti-2004}.  It was found that this model
produces a power-law tail $P(m)\propto1/m^2$ at high $m$.  The reasons
for stability of this law were understood using the Boltzmann kinetic
equation \cite{Yarlagadda-2005,Chakrabarti-2005,Richmond-2005a}, but
most elegantly in the mean-field theory \cite{Mohanty-2006}.  The fat
tail originates from the agents whose saving propensity is close to 1,
who hoard money and do not give it back
\cite{Germano-2005,Patriarca-2006}.  An interesting matrix formulation
of the problem was presented in Ref.\ \cite{Gupta-2006}.  Ref.\
\cite{Patriarca-2007} studied the relaxation rate in the money
transfer models.  Ref.\ \cite{Yakovenko-2000} studied a model with
taxation, which also has an element of proportionality.  The Gamma
distribution was also studied for conservative models within a simple
Boltzmann approach in Ref.\ \cite{Ferrero-2004} and using much more
complicated rules of exchange in Ref.\
\cite{Scafetta-2004a,Scafetta-2004b}.

%%%%%%%%%%%%%%%%%%%%%%%%%%%%%%%%%%%%%%%%%%%%%%%%%%%%%%%%%%%%%%%
\subsection{Additive versus multiplicative models}
\label{Sec:+x}
%%%%%%%%%%%%%%%%%%%%%%%%%%%%%%%%%%%%%%%%%%%%%%%%%%%%%%%%%%%%%%%

The stationary distribution of money (\ref{Gamma}) for the models of
Sec.\ \ref{Sec:saving} is different from the simple exponential
formula (\ref{money}) found for the models of Sec.\ \ref{Sec:BGmoney}.
The origin of this difference can be understood from the Boltzmann
kinetic equation \cite{Wannier,Kinetics}.  This equation describes
time evolution of the distribution function $P(m)$ due to pairwise
interactions:
\begin{eqnarray}
  &&\frac{dP(m)}{dt}=\int\!\!\!\!\int\{
    -f_{[m,m']\to[m-\Delta,m'+\Delta]}P(m)P(m')
\label{Boltzmann}  \\
  &&+f_{[m-\Delta,m'+\Delta]\to[m,m']}
  P(m-\Delta)P(m'+\Delta)\}\,dm'\,d\Delta.
\nonumber
\end{eqnarray}
Here $f_{[m,m']\to[m-\Delta,m'+\Delta]}$ is the probability of
transferring money $\Delta$ from an agent with money $m$ to an agent
with money $m'$ per unit time.  This probability, multiplied by the
occupation numbers $P(m)$ and $P(m')$, gives the rate of transitions
from the state $[m,m']$ to the state $[m-\Delta,m'+\Delta]$.  The
first term in Eq.\ (\ref{Boltzmann}) gives the depopulation rate of
the state $m$.  The second term in Eq.\ (\ref{Boltzmann}) describes
the reversed process, where the occupation number $P(m)$ increases.
When the two terms are equal, the direct and reversed transitions
cancel each other statistically, and the probability distribution is
stationary: $dP(m)/dt=0$.  This is the principle of detailed balance.

In physics, the fundamental microscopic equations of motion of
particles obey time-reversal symmetry.  This means that the
probabilities of the direct and reversed processes are exactly equal:
\begin{eqnarray}
  f_{[m,m']\to[m-\Delta,m'+\Delta]}=f_{[m-\Delta,m'+\Delta]\to[m,m']}.
\label{reversal}
\end{eqnarray}
When Eq.\ (\ref{reversal}) is satisfied, the detailed balance
condition for Eq.\ (\ref{Boltzmann}) reduces to the equation
$P(m)P(m')=P(m-\Delta)P(m'+\Delta)$, because the factors $f$ cancels
out.  The only solution of this equation is the exponential function
$P(m)=c\exp(-m/T_m)$, so the Boltzmann-Gibbs distribution is the
stationary solution of the Boltzmann kinetic equation
(\ref{Boltzmann}).  Notice that the transition probabilities
(\ref{reversal}) are determined by the dynamical rules of the model,
but the equilibrium Boltzmann-Gibbs distribution does not depend on
the dynamical rules at all.  This is the origin of the universality of
the Boltzmann-Gibbs distribution.  It shows that it may be possible to
find out the stationary distribution without knowing details of the
dynamical rules (which are rarely known very well), as long as the
symmetry condition (\ref{reversal}) is satisfied.

The models considered in Sec.\ \ref{Sec:BGmoney} have the
time-reversal symmetry.  The model with the fixed money transfer
$\Delta$ has equal probabilities (\ref{reversal}) of transferring
money from an agent with balance $m$ to an agent with balance $m'$ and
vice versa.  This is also true when $\Delta$ is random, as long as the
probability distribution of $\Delta$ is independent of $m$ and $m'$.
Thus, the stationary distribution $P(m)$ is always exponential in
these models.

However, there is no fundamental reason to expect time-reversal
symmetry in economics, so Eq.\ (\ref{reversal}) may be not valid.  In
this case, the system may have a non-exponential stationary
distribution or no stationary distribution at all.  In model
(\ref{proportional}), the time-reversal symmetry is broken.  Indeed,
when an agent $i$ gives a fixed fraction $\gamma$ of his money $m_i$
to an agent with balance $m_j$, their balances become $(1-\gamma)m_i$
and $m_j+\gamma m_i$.  If we try to reverse this process and appoint
the agent $j$ to be the payer and to give the fraction $\gamma$ of her
money, $\gamma(m_j+\gamma m_i)$, to the agent $i$, the system does not
return to the original configuration $[m_i,m_j]$.  As emphasized by
Angle \cite{Angle-2006}, the payer pays a deterministic fraction of
his money, but the receiver receives a random amount from a random
agent, so their roles are not interchangeable.  Because the
proportional rule typically violates the time-reversal symmetry, the
stationary distribution $P(m)$ in multiplicative models is typically
not exactly exponential.\footnote{However, when $\Delta m$ is a
fraction of the total money $m_i+m_j$ of the two agents, the model is
time-reversible and has an exponential distribution, as discussed in
Sec.\ \ref{Sec:BGmoney}.}  Making the transfer dependent on the money
balance of the payer effectively introduces a Maxwell's demon into the
model.  That is why the stationary distribution is not exponential,
and, thus, does not maximize entropy (\ref{entropy}).  Another view on
the time-reversal symmetry in economic dynamics is presented in Ref.\
\cite{Ao-2007}.

These examples show that the Boltzmann-Gibbs distribution does not
hold for any conservative model.  However, it is universal in a
limited sense.  For a broad class of models that have time-reversal
symmetry, the stationary distribution is exponential and does not
depend on the details of the model.  Conversely, when the
time-reversal symmetry is broken, the distribution may depend on the
details of the model.  The difference between these two classes of
models may be rather subtle.  Deviations from the Boltzmann-Gibbs law
may occur only if the transition rates $f$ in Eq.\ (\ref{reversal})
explicitly depend on the agent's money $m$ or $m'$ in an asymmetric
manner.  Ref.\ \cite{Yakovenko-2000} performed a computer simulation
where the direction of payment was randomly fixed in advance for every
pair of agents $(i,j)$.  In this case, money flows along directed
links between the agents: $i\!\to\!j\!\to\!k$, and the time-reversal
symmetry is strongly violated.  This model is closer to the real
economy, where one typically receives money from an employer and pays
it to a grocery store.  Nevertheless, the Boltzmann-Gibbs distribution
was found in this model, because the transition rates $f$ do not
explicitly depend on $m$ and $m'$ and do not violate Eq.\
(\ref{reversal}).

In the absence of detailed knowledge of real microscopic dynamics of
economic exchanges, the semiuniversal Boltzmann-Gibbs distribution
(\ref{money}) is a natural starting point.  Moreover, the assumption
of Ref.\ \cite{Yakovenko-2000} that agents pay the same prices $\Delta
m$ for the same products, independent of their money balances $m$,
seems very appropriate for the modern anonymous economy, especially
for purchases over the Internet.  There is no particular empirical
evidence for the proportional rules (\ref{proportional}) or
(\ref{saving}).  However, the difference between the additive
(\ref{money}) and multiplicative (\ref{Gamma}) distributions may be
not so crucial after all.  From the mathematical point of view, the
difference is in the implementation of the boundary condition at
$m=0$.  In the additive models of Sec.\ \ref{Sec:BGmoney}, there is a
sharp cut-off of $P(m)\neq0$ at $m=0$.  In the multiplicative models
of Sec.\ \ref{Sec:saving}, the balance of an agent never reaches
$m=0$, so $P(m)$ vanishes at $m\to0$ in a power-law manner.  At the
same time, $P(m)$ decreases exponentially for large $m$ for both
models.

By further modifying the rules of money transfer and introducing more
parameters in the models, one can obtain even more complicated
distributions \cite{Scafetta-2007}.  However, one can argue that
parsimony is the virtue of a good mathematical model, not the
abundance of additional assumptions and parameters, whose
correspondence to reality is hard to verify.

%%%%%%%%%%%%%%%%%%%%%%%%%%%%%%%%%%%%%%%%%%%%%%%%%%%%%%%%%%%%%%%
\section{Statistical Mechanics of Wealth Distribution}
\label{Sec:wealth}
%%%%%%%%%%%%%%%%%%%%%%%%%%%%%%%%%%%%%%%%%%%%%%%%%%%%%%%%%%%%%%%

In the econophysics literature on exchange models, the terms ``money''
and ``wealth'' are often used interchangeably; however, economists
emphasize the difference between these two concepts.  In this section,
we review the models of wealth distribution, as opposed to money
distribution.

%%%%%%%%%%%%%%%%%%%%%%%%%%%%%%%%%%%%%%%%%%%%%%%%%%%%%%%%%%%%%%%
\subsection{Models with a conserved commodity}
\label{Sec:w=const}
%%%%%%%%%%%%%%%%%%%%%%%%%%%%%%%%%%%%%%%%%%%%%%%%%%%%%%%%%%%%%%%

What is the difference between money and wealth?  On can argue
\cite{Yakovenko-2000} that wealth $w_i$ is equal to money $m_i$ plus
the other property that an agent $i$ has.  The latter may include
durable material property, such as houses and cars, and financial
instruments, such as stocks, bonds, and options.  Money (paper cash,
bank accounts) is generally liquid and countable.  However, the other
property is not immediately liquid and has to be sold first (converted
into money) to be used for other purchases.  In order to estimate the
monetary value of property, one needs to know the price $p$.  In the
simplest model, let us consider just one type of property, say, stocks
$s$.  Then the wealth of an agent $i$ is given by the formula
\begin{equation}
  w_i=m_i+p\,s_i.
\label{wealth}
\end{equation}
It is assumed that the price $p$ is common for all agents and is
established by some kind of market process, such as an auction, and
may change in time.

It is reasonable to start with a model where both the total money
$M=\sum_im_i$ and the total stock $S=\sum_is_i$ are conserved
\cite{Chakrabarti-2001,Chakrabarti-2006,Ausloos-2007}.  The agents pay
money to buy stock and sell stock to get money, and so on.  Although
$M$ and $S$ are conserved, the total wealth $W=\sum_iw_i$ is generally
not conserved, because of price fluctuation \cite{Chakrabarti-2006} in
Eq.\ (\ref{wealth}).  This is an important difference from the money
transfers models of Sec.\ \ref{Sec:money}.  Here the wealth $w_i$ of
an agent $i$, not participating in any transactions, may change when
transactions between other agents establish a new price $p$.
Moreover, the wealth $w_i$ of an agent $i$ does not change after a
transaction with an agent $j$.  Indeed, in exchange for paying money
$\Delta m$, agent $i$ receives the stock $\Delta s=\Delta m/p$, so her
total wealth (\ref{wealth}) remains the same.  In principle, the agent
can instantaneously sell the stock back at the same price and recover
the money paid.  If the price $p$ never changes, then the wealth $w_i$
of each agent remains constant, despite transfers of money and stock
between agents.

We see that redistribution of wealth in this model is directly related
to price fluctuations.  One mathematical model of this process was
studied in Ref.\ \cite{Silver-2002}.  In this model, the agents
randomly change preferences for the fraction of their wealth invested
in stocks.  As a result, some agents offer stock for sale and some
want to buy it.  The price $p$ is determined from the market-clearing
auction matching supply and demand.  Ref.\ \cite{Silver-2002}
demonstrated in computer simulations and proved analytically using the
theory of Markov processes that the stationary distribution $P(w)$ of
wealth $w$ in this model is given by the Gamma distribution, as in
Eq.\ (\ref{Gamma}).  Various modifications of this model
\cite{Lux-2005}, such as introducing monopolistic coalitions, do not
change this result significantly, which shows the robustness of the
Gamma distribution.  For models with a conserved commodity, Ref.\
\cite{Chakrabarti-2006} found the Gamma distribution for a fixed
saving propensity and a power law tail for a distributed saving
propensity.

Another model with conserved money and stock was studied in Ref.\
\cite{Raberto-2003} for an artificial stock market, where traders
follow different investment strategies: random, momentum, contrarian,
and fundamentalist.  Wealth distribution in the model with random
traders was found have a power-law tail $P(w)\sim1/w^2$ for large $w$.
However, unlike in most other simulation, where all agents initially
have equal balances, here the initial money and stock balances of the
agents were randomly populated according to a power law with the same
exponent.  This raises the question whether the observed power-law
distribution of wealth is an artifact of the initial conditions,
because equilibrization of the upper tail may take a very long
simulation time.

%%%%%%%%%%%%%%%%%%%%%%%%%%%%%%%%%%%%%%%%%%%%%%%%%%%%%%%%%%%%%%%
\subsection{Models with stochastic growth of wealth}
\label{Sec:w-not-const}
%%%%%%%%%%%%%%%%%%%%%%%%%%%%%%%%%%%%%%%%%%%%%%%%%%%%%%%%%%%%%%%

Although the total wealth $W$ is not exactly conserved in the models
considered in Sec.\ \ref{Sec:w=const}, nevertheless $W$ remains
constant on average, because the total money $M$ and stock $S$ are
conserved.  A different model for wealth distribution was proposed in
Ref.\ \cite{Bouchaud-2000}.  In this model, time evolution of the
wealth $w_i$ of an agent $i$ is given by the stochastic differential
equation
\begin{equation}
  \frac{dw_i}{dt}=\eta_i(t)\,w_i + \sum_{j(\neq i)} J_{ij}w_j -
  \sum_{j(\neq i)} J_{ji}w_i,
\label{Bouchaud}
\end{equation}
where $\eta_i(t)$ is a Gaussian random variable with the mean
$\langle\eta\rangle$ and the variance $2\sigma^2$.  This variable
represents growth or loss of wealth of an agent due to investment in
stock market.  The last two terms describe transfer of wealth between
different agents, which is taken to be proportional to the wealth of
the payers with the coefficients $J_{ij}$.  So, the model
(\ref{Bouchaud}) is multiplicative and invariant under the scale
transformation $w_i\to Zw_i$.  For simplicity, the exchange fractions
are taken to be the same for all agents: $J_{ij}=J/N$ for all $i\neq
j$, where $N$ is the total number of agents.  In this case, the last
two terms in Eq.\ (\ref{Bouchaud}) can be written as $J(\langle
w\rangle - w_i)$, where $\langle w\rangle=\sum_iw_i/N$ is the average
wealth per agent.  This case represents a ``mean-field'' model, where
all agents feel the same environment.  It can be easily shown that the
average wealth increases in time as $\langle w\rangle_t=\langle
w\rangle_0e^{(\langle\eta\rangle+\sigma^2)t}$.  Then, it makes more
sense to consider the relative wealth $\tilde w_i=w_i/\langle
w\rangle_t$.  Eq.\ (\ref{Bouchaud}) for this variable becomes
\begin{equation}
  \frac{d\tilde
  w_i}{dt}=(\eta_i(t)-\langle\eta\rangle-\sigma^2)\,\tilde w_i
  +J(1-\tilde w_i).
\label{relative}
\end{equation}
The probability distribution $P(\tilde w,t)$ for the stochastic
differential equation (\ref{relative}) is governed by the
Fokker-Planck equation
\begin{equation}
  \frac{\partial P}{\partial t}
  =\frac{\partial[J(\tilde w-1)+\sigma^2\tilde w] P}{\partial \tilde w}
  +\sigma^2\frac{\partial}{\partial\tilde w}
  \left(\tilde w\frac{\partial(\tilde wP)}{\partial\tilde w}\right).
\label{FP-Bouchaud}
\end{equation}
The stationary solution ($\partial P/\partial t=0$) of this equation
is given by the following formula
\begin{equation}
  P(\tilde w)=c\,\frac{e^{-J/\sigma^2\tilde w}}
  {\tilde w^{2+J/\sigma^2}}.
\label{P-Bouchaud}
\end{equation}
The distribution (\ref{P-Bouchaud}) is quite different from the
Boltzmann-Gibbs (\ref{money}) and Gamma (\ref{Gamma}) distributions.
Eq.\ (\ref{P-Bouchaud}) has a power-law tail at large $\tilde w$ and a
sharp cutoff at small $\tilde w$.  Eq.\ (\ref{Bouchaud}) is a version
of the generalized Lotka-Volterra model, and the stationary
distribution (\ref{P-Bouchaud}) was also obtained in Ref.\
\cite{Solomon-2001,Solomon-2002}.  The model was generalized to
include negative wealth in Ref.\ \cite{Huang-2004}.

Ref.\ \cite{Bouchaud-2000} used the mean-field approach.  A similar
result was found for a model with pairwise interaction between agents
in Ref.\ \cite{Slanina-2004}.  In this model, wealth is transferred
between the agents using the proportional rule (\ref{proportional}).
In addition, the wealth of the agents increases by the factor
$1+\zeta$ in each transaction.  This factor is supposed to reflect
creation of wealth in economic interactions. Because the total wealth
in the system increases, it makes sense to consider the distribution
of relative wealth $P(\tilde w)$.  In the limit of continuous trading,
Ref.\ \cite{Slanina-2004} found the same stationary distribution
(\ref{P-Bouchaud}).  This result was reproduced using a mathematically
more involved treatment of this model in Ref.\ \cite{Toscani-2005}.
Numerical simulations of the models with stochastic noise $\eta$ in
Ref.\ \cite{Scafetta-2004a,Scafetta-2004b} also found a power law tail
for large $w$.

Let us contrast the models discussed in Secs.\ \ref{Sec:w=const} and
\ref{Sec:w-not-const}.  In the former case, where money and commodity
are conserved, and wealth does not grow, the distribution of wealth is
given by the Gamma distribution with the exponential tail for large
$w$.  In the latter models, wealth grows in time exponentially, and
the distribution of relative wealth has a power law tail for large
$\tilde w$.  These results suggest that the presence of a power-law
tail is a nonequilibrium effect that requires constant growth or
inflation of the economy, but disappears for a closed system with
conservation laws.

Reviews of the discussed models were also given in Refs.\
\cite{Richmond-2006a,Richmond-2006b}.  Because of lack of space, we
omit discussion of models with wealth condensation
\cite{Bouchaud-2000,Redner-1998,Burda-2002,Iglesias-2003a,Braun-2006},
where few agents accumulate a finite fraction of total wealth, and
studies of wealth distribution on networks
\cite{Coelho-2005,Iglesias-2003b,DiMatteo-2004,Hu-2007}.  In this
section, we discussed the models with long-range interaction, where
any agent can exchange money and wealth with any other agent.  A local
model, where agents trade only with the nearest neighbors, was studied
in Ref.\ \cite{Bak-1999}.

%%%%%%%%%%%%%%%%%%%%%%%%%%%%%%%%%%%%%%%%%%%%%%%%%%%%%%%%%%%%%%%
\subsection{Empirical data on money and wealth distributions}
\label{Sec:w-empirical}
%%%%%%%%%%%%%%%%%%%%%%%%%%%%%%%%%%%%%%%%%%%%%%%%%%%%%%%%%%%%%%%

\begin{figure}[b]
\includegraphics[width=0.9\linewidth]{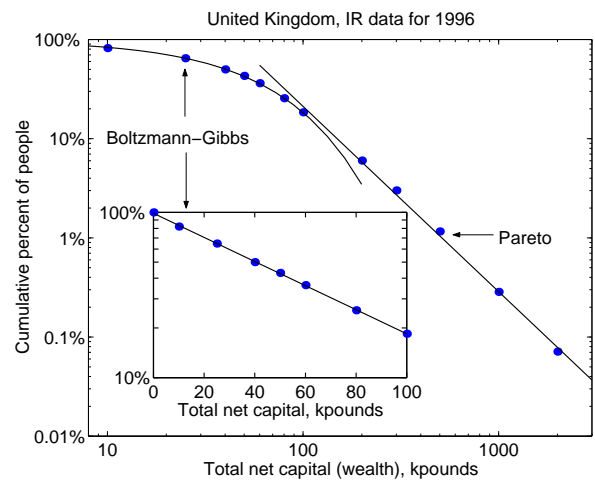}
\caption{Cumulative probability distribution of net wealth in the UK
  shown on log-log (main panel) and log-linear (inset) scales.  Points
  represent the data from the Inland Revenue, and solid lines are fits
  to the exponential (Boltzmann-Gibbs) and power (Pareto) laws.
  (Reproduced from Ref.\ \cite{Yakovenko-2001b})}
\label{Fig:UKwealth}
\end{figure}

It would be very interesting to compare theoretical results for money
and wealth distributions in various models with empirical data.
Unfortunately, such empirical data are difficult to find.  Unlike
income, which is discussed in Sec.\ \ref{Sec:income}, wealth is not
routinely reported by the majority of individuals to the government.
However, in many countries, when a person dies, all assets must be
reported for the purpose of inheritance tax.  So, in principle, there
exist good statistics of wealth distribution among dead people, which,
of course, is different from the wealth distribution among living
people.  Using an adjustment procedure based on the age, gender, and
other characteristics of the deceased, the UK tax agency, the Inland
Revenue, reconstructed the wealth distribution of the whole population
of the UK \cite{UKwealth}.  Fig.\ \ref{Fig:UKwealth} shows the UK data
for 1996 reproduced from Ref.\ \cite{Yakovenko-2001b}.  The figure
shows the cumulative probability $C(w)=\int_w^\infty P(w')\,dw'$ as a
function of the personal net wealth $w$, which is composed of assets
(cash, stocks, property, household goods, etc.)\ and liabilities
(mortgages and other debts).  Because statistical data are usually
reported at non-uniform intervals of $w$, it is more practical to plot
the cumulative probability distribution $C(w)$ rather than its
derivative, the probability density $P(w)$.  Fortunately, when $P(w)$
is an exponential or a power-law function, then $C(w)$ is also an
exponential or a power-law function.

%%%%%%%%%%%%%%%%%%%%%%%%%%%%%%%%%%%%%%%%%%%%%%%%%%%%%%%%%%%%%%%
%%%%%%%%%%%%%%%%%%%%%%%%%%%%%%%%%%%%%%%%%%%%%%%%%%%%%%%%%%%%%%%

The main panel in Fig.\ \ref{Fig:UKwealth} shows a plot of $C(w)$ on a
log-log scale, where a straight line represents a power-law
dependence.  The figure shows that the distribution follows a power
law $C(w)\propto1/w^\alpha$ with exponent $\alpha=1.9$ for the wealth
greater than about 100~k$\pounds$.  The inset in Fig.\
\ref{Fig:UKwealth} shows the data on log-linear scale, where a
straight line represents an exponential dependence.  We observe that,
below 100~k$\pounds$, the data is well fitted by the exponential
distribution $C(w)\propto\exp(-w/T_w)$ with the effective ``wealth
temperature'' $T_w=60$~k$\pounds$ (which corresponds to the median
wealth of 41~k$\pounds$).  So, the distribution of wealth is
characterized by the Pareto power law in the upper tail of the
distribution and the exponential Boltzmann-Gibbs law in the lower part
of the distribution for the great majority (about 90\%) of the
population.  Similar results are found for the distribution of income,
as discussed in Sec.\ \ref{Sec:income}.  One may speculate that wealth
distribution in the lower part is dominated by distribution of money,
because the corresponding people do not have other significant assets,
so the results of Sec.\ \ref{Sec:money} give the Boltzmann-Gibbs law.
On the other hand, the upper tail of wealth distribution is dominated
by investment assess, where the results of Sec.\ \ref{Sec:w-not-const}
give the Pareto law.  The power law was studied by many researchers
for the upper-tail data, such as the Forbes list of 400 richest people
\cite{Solomon-2007,Sinha-2006}, but much less attention was paid to
the lower part of the wealth distribution.  Curiously, Ref.\
\cite{Abul-Magd-2002} found that the wealth distribution in the
ancient Egyptian society was consistent with Eq.\ (\ref{P-Bouchaud}).

For direct comparison with the results of Sec.\ \ref{Sec:money}, it
would be very interesting to find data on the distribution of money,
as opposed to the distribution of wealth.  Making a reasonable
assumption that most people keep most of their money in banks, one can
approximate the distribution of money by the distribution of balances
on bank accounts.  (Balances on all types of bank accounts, such as
checking, saving, and money manager, associated with the same person
should be added up.)  Despite imperfections (people may have accounts
in different banks or not keep all their money in banks), the
distribution of balances on bank accounts would give valuable
information about the distribution of money.  The data for a big
enough bank would be representative of the distribution in the whole
economy.  Unfortunately, it has not been possible to obtain such data
thus far, even though it would be completely anonymous and not
compromise privacy of bank clients.

Measuring the probability distribution of money would be very useful,
because it determines how much people can, in principle, spend on
purchases without going into debt.  This is different from the
distribution of wealth, where the property component, such as house,
car, or retirement investment, is effectively locked up and, in most
cases, is not easily available for consumer spending.  So, although
wealth distribution may reflect the distribution of economic power,
the distribution of money is more relevant for consumption.  Money
distribution can be useful for determining prices that maximize
revenue of a manufacturer \cite{Yakovenko-2000}.  If a price $p$ is
set too high, few people can afford it, and, if a price is too low,
the sales revenue is small, so the optimal price must be in between.
The fraction of population who can afford to pay the price $p$ is
given by the cumulative probability $C(p)$, so the total sales revenue
is proportional to $pC(p)$.  For the exponential distribution
$C(p)=\exp(-p/T_m)$, the maximal revenue is achieved at $p=T_m$,
i.e.,\ the optimal price is equal to the average amount of money per
person \cite{Yakovenko-2000}.  Indeed, the prices of mass-market
consumer products, such as computers, electronics, and appliances,
remain stable for many years at a level determined by their
affordability to the population, whereas technical parameters of these
products at the same price level improve dramatically owing to
technological progress.

%%%%%%%%%%%%%%%%%%%%%%%%%%%%%%%%%%%%%%%%%%%%%%%%%%%%%%%%%%%%%%%
\section{Data and Models for Income Distribution}
\label{Sec:income}
%%%%%%%%%%%%%%%%%%%%%%%%%%%%%%%%%%%%%%%%%%%%%%%%%%%%%%%%%%%%%%%

In contrast to money and wealth distributions, a lot more empirical
data are available for the distribution of income $r$ from tax
agencies and population surveys.  In this section, we first present
empirical data on income distribution and then discuss theoretical
models.

%%%%%%%%%%%%%%%%%%%%%%%%%%%%%%%%%%%%%%%%%%%%%%%%%%%%%%%%%%%%%%%
\subsection{Empirical data on income distribution}
\label{Sec:r-data}
%%%%%%%%%%%%%%%%%%%%%%%%%%%%%%%%%%%%%%%%%%%%%%%%%%%%%%%%%%%%%%%

\begin{figure}[b]
\includegraphics[width=0.9\linewidth]{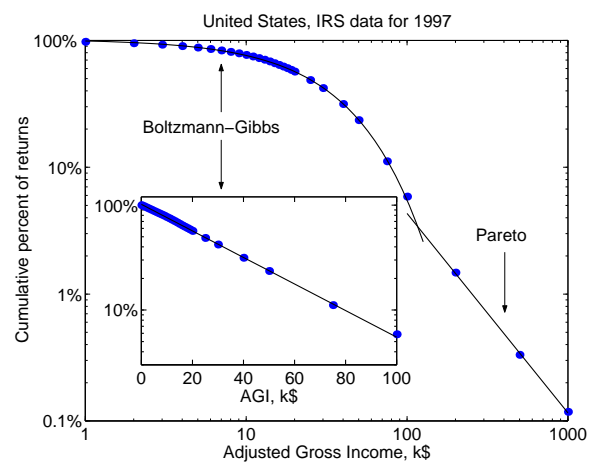}
\caption{Cumulative probability distribution of tax returns for USA in
  1997 shown on log-log (main panel) and log-linear (inset) scales.
  Points represent the Internal Revenue Service data, and solid lines
  are fits to the exponential and power-law functions.  (Reproduced
  from Ref.\ \cite{Yakovenko-2003})}
\label{Fig:income1997}
\end{figure}

Empirical studies of income distribution have a long history in the
economic literature \cite{Kakwani,Cowell,Atkinson}.  Following the
work by Pareto \cite{Pareto}, much attention was focused on the
power-law upper tail of the income distribution and less on the lower
part.  In contrast to more complicated functions discussed in
literature, Ref.\ \cite{Yakovenko-2001a} introduced a new idea by
demonstrating that the lower part of income distribution can be well
fitted with a simple exponential function $P(r)=c\exp(-r/T_r)$
characterized by just one parameter, the ``income temperature'' $T_r$.
Then it was recognized that the whole income distribution can be
fitted by an exponential function in the lower part and a power-law
function in the upper part \cite{Yakovenko-2001b,Yakovenko-2003}, as
shown in Fig.\ \ref{Fig:income1997}.  The straight line on the
log-linear scale in the inset of Fig.\ \ref{Fig:income1997}
demonstrates the exponential Boltzmann-Gibbs law, and the straight
line on the log-log scale in the main panel illustrates the Pareto
power law.  The fact that income distribution consists of two distinct
parts reveals the two-class structure of the American society
\cite{Yakovenko-2005b,Yakovenko-2005a}.  Coexistence of the
exponential and power-law distributions is also known in plasma
physics and astrophysics, where they are called the ``thermal'' and
``superthermal'' parts \cite{Hasegawa-1985,Mason-2003,Collier-2004}.
The boundary between the lower and upper classes can be defined as the
intersection point of the exponential and power-law fits in Fig.\
\ref{Fig:income1997}.  For 1997, the annual income separating the two
classes was about 120~k\$.  About 3\% of the population belonged to
the upper class, and 97\% belonged to the lower class.

%%%%%%%%%%%%%%%%%%%%%%%%%%%%%%%%%%%%%%%%%%%%%%%%%%%%%%%%%%%%%%%
%%%%%%%%%%%%%%%%%%%%%%%%%%%%%%%%%%%%%%%%%%%%%%%%%%%%%%%%%%%%%%%

\begin{figure}[b]
\includegraphics[angle=-90,width=0.95\linewidth]{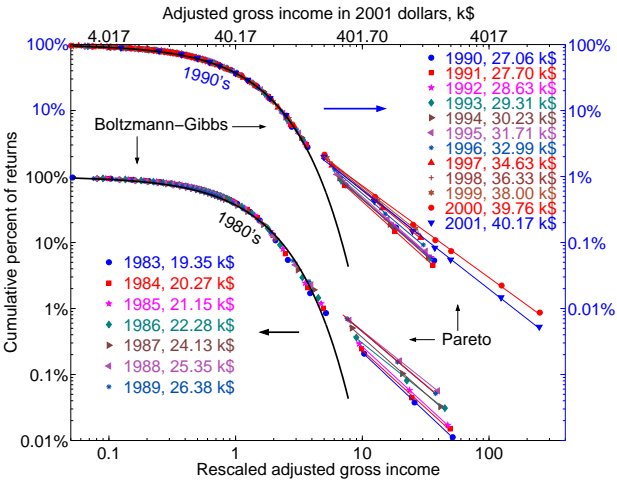}
\caption{Cumulative probability distribution of tax returns plotted on
  log-log scale versus $r/T_r$ (the annual income $r$ normalized by
  the average income $T_r$ in the exponential part of the
  distribution).  The IRS data points are for 1983--2001, and the
  columns of numbers give the values of $T_r$ for the corresponding
  years.  (Reproduced from Ref.\ \cite{Yakovenko-2005a})}
\label{Fig:income}
\end{figure}

Ref.\ \cite{Yakovenko-2005a} studied time evolution of income
distribution in the USA during 1983--2001 on the basis of the data
from the Internal Revenue Service (IRS), the government tax agency.
The structure of the income distribution was found to be qualitatively
the same for all years, as shown in Fig.\ \ref{Fig:income}.  The
average income in nominal dollars approximately doubled during this
time interval.  So, the horizontal axis in Fig.\ \ref{Fig:income} shows
the normalized income $r/T_r$, where the ``income temperature'' $T_r$
was obtained by fitting of the exponential part of the distribution
for each year.  The values of $T_r$ are shown in Fig.\
\ref{Fig:income}.  The plots for the 1980s and 1990s are shifted
vertically for clarity.  We observe that the data points in the
lower-income part of the distribution collapse on the same exponential
curve for all years.  This demonstrates that the shape of the income
distribution for the lower class is extremely stable and does not
change in time, despite gradual increase of the average income in
nominal dollars.  This observation suggests that the lower-class
distribution is in statistical, ``thermal'' equilibrium.

On the other hand, Fig.\ \ref{Fig:income} shows that the income
distribution in the upper class does not rescale and significantly
changes in time.  Ref.\ \cite{Yakovenko-2005a} found that the exponent
$\alpha$ of the power law $C(r)\propto1/r^\alpha$ decreased from 1.8
in 1983 to 1.4 in 2000.  This means that the upper tail became
``fatter''.  Another useful parameter is the total income of the upper
class as the fraction $f$ of the total income in the system.  The
fraction $f$ increased from 4\% in 1983 to 20\% in 2000
\cite{Yakovenko-2005a}.  However, in year 2001, $\alpha$ increased and
$f$ decreases, indicating that the upper tail was reduced after the
stock market crash at that time.  These results indicate that the
upper tail is highly dynamical and not stationary.  It tends to swell
during the stock market boom and shrink during the bust.  Similar
results were found for Japan
\cite{Souma-2001,Souma-2002,Aoki-2003a,Aoki-2003b}.

%%%%%%%%%%%%%%%%%%%%%%%%%%%%%%%%%%%%%%%%%%%%%%%%%%%%%%%%%%%%%%%%%%%
%%%%%%%%%%%%%%%%%%%%%%%%%%%%%%%%%%%%%%%%%%%%%%%%%%%%%%%%%%%%%%%%%%%%%%

\begin{figure}[b]
\includegraphics[angle=-90,width=0.78\linewidth]{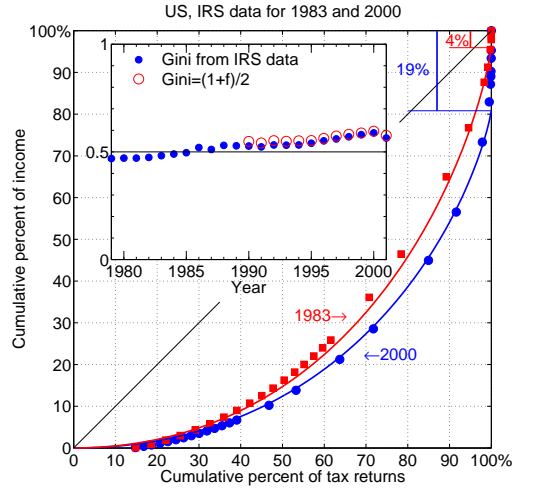}
\caption{\textit{Main panel:} Lorenz plots for income distribution in
  1983 and 2000.  The data points are from the IRS \cite{Petska-2003},
  and the theoretical curves represent Eq.\ (\ref{Lorenz}) with $f$
  from Fig.\ \ref{Fig:income}.  \textit{Inset:} The closed circles are
  the IRS data \cite{Petska-2003} for the Gini coefficient $G$, and
  the open circles show the theoretical formula $G=(1+f)/2$.
  (Reproduced from Ref.\ \cite{Yakovenko-2005a})}
\label{Fig:Lorenz}
\end{figure}

Although relative income inequality within the lower class remains
stable, the overall income inequality in the USA has increased
significantly as a result of the tremendous growth of the income of
the upper class.  This is illustrated by the Lorenz curve and the Gini
coefficient shown in Fig.\ \ref{Fig:Lorenz}.  The Lorenz curve
\cite{Kakwani} is a standard way of representing income distribution
in the economic literature.  It is defined in terms of two coordinates
$x(r)$ and $y(r)$ depending on a parameter $r$:
\begin{equation}
  x(r)=\int_0^r P(r')\,dr',\quad
  y(r)=\frac{\int_0^r r' P(r')\,dr'}{\int_0^\infty r' P(r')\,dr'}.
\label{xy}
\end{equation}
The horizontal coordinate $x(r)$ is the fraction of the population
with income below $r$, and the vertical coordinate $y(r)$ is the
fraction of the income this population accounts for.  As $r$ changes
from 0 to $\infty$, $x$ and $y$ change from 0 to 1 and parametrically
defines a curve in the $(x,y)$-plane.

Fig.\ \ref{Fig:Lorenz} shows the data points for the Lorenz curves in
1983 and 2000, as computed by the IRS \cite{Petska-2003}.  Ref.\
\cite{Yakovenko-2001a} analytically derived the Lorenz curve formula
$y=x+(1-x)\ln(1-x)$ for a purely exponential distribution
$P(r)=c\exp(-r/T_r)$.  This formula is shown by the red line in Fig.\
\ref{Fig:Lorenz} and describes the 1983 data reasonably well.
However, for year 2000, it is essential to take into account the
fraction $f$ of income in the upper tail, which modifies for the
Lorenz formula as follows
\cite{Yakovenko-2003,Yakovenko-2005b,Yakovenko-2005a}
\begin{equation}
  y=(1-f)[x+(1-x)\ln(1-x)]+f\,\Theta(x-1).
\label{Lorenz}
\end{equation}
The last term in Eq.\ (\ref{Lorenz}) represent the vertical jump of
the Lorenz curve at $x=1$, where a very small percentage of population
in the upper class accounts for a substantial fraction $f$ of the
total income.  The blue curve representing Eq.\ (\ref{Lorenz}) fits
the 2000 data in Fig.\ \ref{Fig:Lorenz} very well.

%%%%%%%%%%%%%%%%%%%%%%%%%%%%%%%%%%%%%%%%%%%%%%%%%%%%%%%%%%%%%%%%%%
%%%%%%%%%%%%%%%%%%%%%%%%%%%%%%%%%%%%%%%%%%%%%%%%%%%%%%%%%%%%%%%%%%

The deviation of the Lorenz curve from the straight diagonal line in
Fig.\ \ref{Fig:Lorenz} is a certain measure of income inequality.
Indeed, if everybody had the same income, the Lorenz curve would be
a diagonal line, because the fraction of income would be proportional
to the fraction of the population.  The standard measure of income
inequality is the so-called Gini coefficient $0\leq G\leq1$, which is
defined as the area between the Lorenz curve and the diagonal line,
divided by the area of the triangle beneath the diagonal line
\cite{Kakwani}.  Time evolution of the Gini coefficient, as computed
by the IRS \cite{Petska-2003}, is shown in the inset of Fig.\
\ref{Fig:Lorenz}.  Ref.\ \cite{Yakovenko-2001a} derived analytically
the result that $G=1/2$ for a purely exponential distribution.  In the
first approximation, the values of $G$ shown in the inset of Fig.\
\ref{Fig:Lorenz} are indeed close to the theoretical value 1/2.  If we
take into account the upper tail using Eq.\ (\ref{Lorenz}), the
formula for the Gini coefficient becomes $G=(1+f)/2$
\cite{Yakovenko-2005a}.  The inset in Fig.\ \ref{Fig:Lorenz} shows
that this formula gives a very good fit to the IRS data for the 1990s
using the values of $f$ deduced from Fig.\ \ref{Fig:income}.  The
values $G<1/2$ in the 1980s cannot be captured by this formula,
because the Lorenz data points are slightly above the theoretical
curve for 1983 in Fig.\ \ref{Fig:Lorenz}.  Overall, we observe that
income inequality has been increasing for the last 20 years, because
of swelling of the Pareto tail, but decreased in 2001 after the stock
market crash.

%%%%%%%%%%%%%%%%%%%%%%%%%%%%%%%%%%%%%%%%%%%%%%%%%%%%%%%%%%%%%%%%%%
\begin{figure}[t]
\includegraphics[width=0.9\linewidth]{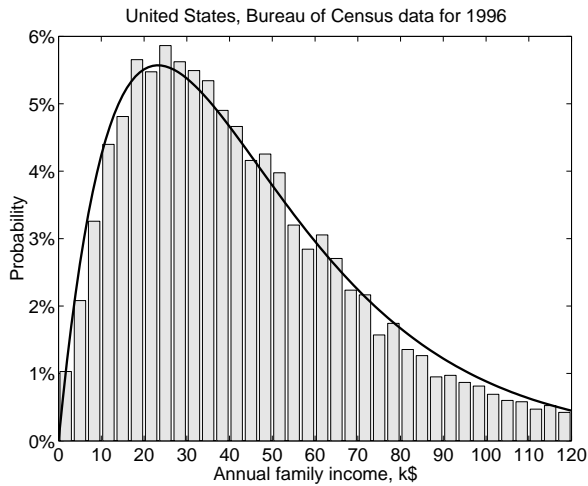}
\caption{\textit{Histogram:} Probability distribution of family income
  for families with two adults (US Census Bureau data).  \textit{Solid
  line:} Fit to Eq.\ (\ref{family}).  (Reproduced from Ref.\
  \cite{Yakovenko-2001a}.)}
\label{Fig:income-2}
\end{figure}
%%%%%%%%%%%%%%%%%%%%%%%%%%%%%%%%%%%%%%%%%%%%%%%%%%%%%%%%%%%%%%%%%%

%%%%%%%%%%%%%%%%%%%%%%%%%%%%%%%%%%%%%%%%%%%%%%%%%%%%%%%%%%%%%%%%%%
\begin{figure}[b]
\includegraphics[width=0.8\linewidth]{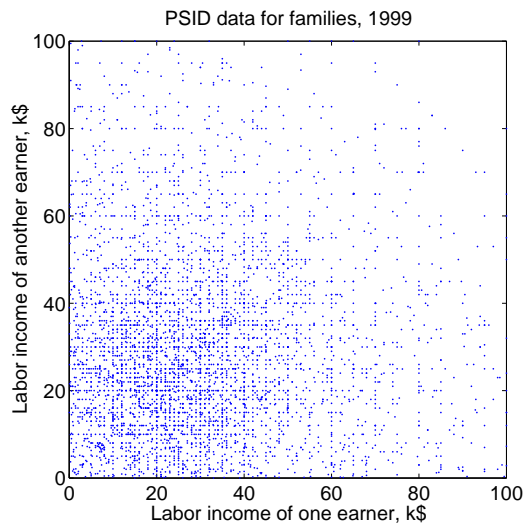}
\caption{Scatter plot of the spouses' incomes $(r_1,r_2)$ and
  $(r_2,r_1)$ based on the data from the Panel Study of Income
  Dynamics (PSID).  (Reproduced from Ref.\ \cite{Yakovenko-2003})}
\label{Fig:scatter}
\end{figure}
%%%%%%%%%%%%%%%%%%%%%%%%%%%%%%%%%%%%%%%%%%%%%%%%%%%%%%%%%%%%%%%%%%

Thus far we discussed the distribution of individual income.  An
interesting related question is the distribution $P_2(r)$ of family
income $r=r_1+r_2$, where $r_1$ and $r_2$ are the incomes of spouses.
If individual incomes are distributed exponentially
$P(r)\propto\exp(-r/T_r)$, then
\begin{equation}
  P_2(r)=\int_0^r dr' P(r')P(r-r')=c\,r\exp(-r/T_r),
\label{family}
\end{equation}
where $c$ is a normalization constant.  Fig.\ \ref{Fig:income-2} shows
that Eq.\ (\ref{family}) is in good agreement with the family income
distribution data from the US Census Bureau \cite{Yakovenko-2001a}.
In Eq.\ (\ref{family}), we assumed that incomes of spouses are
uncorrelated.  This simple approximation is indeed supported by the
scatter plot of incomes of spouses shown in Fig.\ \ref{Fig:scatter}.
Each family is represented in this plot by two points $(r_1,r_2)$ and
$(r_2,r_1)$ for symmetry.  We observe that the density of points is
approximately constant along the lines of constant family income
$r_1+r_2=\rm const$, which indicates that incomes of spouses are
approximately uncorrelated.  There is no significant clustering of
points along the diagonal $r_1=r_2$, i.e.,\ no strong positive
correlation of spouses' incomes.

\begin{figure}
\includegraphics[width=0.78\linewidth]{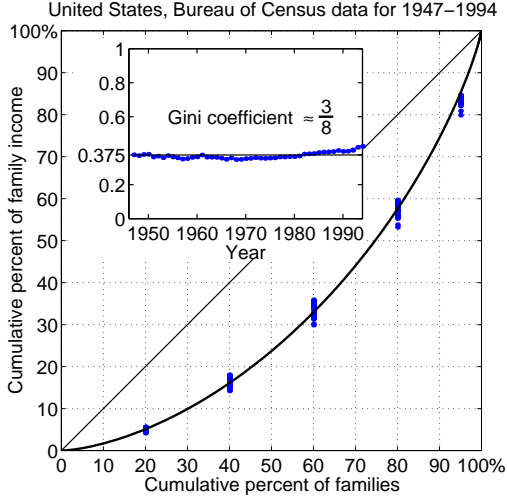}
\caption{\textit{Main panel:} Lorenz plot for family income calculated
  from Eq.\ (\ref{family}), compared with the US Census data points.
  \textit{Inset:} The US Census data points for the Gini coefficient
  for families, compared with the theoretically calculated value
  3/8=37.5\%.  (Reproduced from Ref.\ \cite{Yakovenko-2001a})}
\label{Fig:Lorenz-2}
\end{figure}

The Gini coefficient for the family income distribution (\ref{family})
was calculated in Ref.\ \cite{Yakovenko-2001a} as $G=3/8=37.5\%$.
Fig.\ \ref{Fig:Lorenz-2} shows the Lorenz quintiles and the Gini
coefficient for 1947--1994 plotted from the US Census Bureau data.
The solid line, representing the Lorenz curve calculated from Eq.\
(\ref{family}), is in good agreement with the data.  The systematic
deviation for the top 5\% of earners results from the upper tail,
which has a less pronounced effect on family income than on individual
income, because of income averaging in the family.  The Gini
coefficient, shown in the inset of Fig.\ \ref{Fig:Lorenz-2}, is close
to the calculated value of $37.5\%$.  Moreover, the average $G$ for
the developed capitalist countries of North America and western
Europe, as determined by the World Bank \cite{Yakovenko-2003}, is also
close to the calculated value 37.5\%.

%%%%%%%%%%%%%%%%%%%%%%%%%%%%%%%%%%%%%%%%%%%%%%%%%%%%%%%%%%%%%%%%%%
%%%%%%%%%%%%%%%%%%%%%%%%%%%%%%%%%%%%%%%%%%%%%%%%%%%%%%%%%%%%%%%%%%

Income distribution has been examined in econophysics papers for
different countries: Japan
\cite{Souma-2001,Souma-2002,Aoki-2003a,Aoki-2003b,Souma-2005,Nirei-2007,Ferrero-2004,Ferrero-2005},
Germany \cite{Clementi-2005a,Clementi-2007}, the UK
\cite{Richmond-2006b,Ferrero-2004,Ferrero-2005,Clementi-2005a,Clementi-2007},
Italy \cite{Clementi-2005b,Clementi-2006,Clementi-2007}, the USA
\cite{Clementi-2005a,Rawlings-2004}, India \cite{Sinha-2006},
Australia \cite{DiMatteo-2004,Clementi-2006,Yakovenko-2006}, and New
Zealand \cite{Ferrero-2004,Ferrero-2005}.  The distributions are
qualitatively similar to the results presented in this section.  The
upper tail follows a power law and comprises a small fraction of
population.  To fit the lower part of the distribution, different
papers used exponential, Gamma, and log-normal distributions.
Unfortunately, income distribution is often reported by statistical
agencies for households, so it is difficult to differentiate between
one-earner and two-earner income distributions.  Some papers used
interpolating functions with different asymptotic behavior for low and
high incomes, such as the Tsallis function \cite{Ferrero-2005} and the
Kaniadakis function \cite{Clementi-2007}.  However, the transition
between the lower and upper classes is not smooth for the US data
shown in Figs.\ \ref{Fig:income1997} and \ref{Fig:income}, so such
functions would not be useful in this case.  The special case is
income distribution in Argentina during the economic crisis, which
shows a time-dependent bimodal shape with two peaks
\cite{Ferrero-2005}.

%%%%%%%%%%%%%%%%%%%%%%%%%%%%%%%%%%%%%%%%%%%%%%%%%%%%%%%%%%%%%%%
\subsection{Theoretical models of income distribution}
\label{Sec:r-theory}
%%%%%%%%%%%%%%%%%%%%%%%%%%%%%%%%%%%%%%%%%%%%%%%%%%%%%%%%%%%%%%%

Having examined the empirical data on income distribution, let us now
discuss theoretical models.  Income $r_i$ is the influx of money per
unit time to an agent $i$.  If the money balance $m_i$ is analogous to
energy, then the income $r_i$ would be analogous to power, which is
the energy flux per unit time.  So, one should conceptually
distinguish between the distributions of money and income.  While
money is regularly transferred from one agent to another in pairwise
transactions, it is not typical for agents to trade portions of their
income.  Nevertheless, indirect transfer of income may occur when one
employee is promoted and another demoted while the total annual budget
is fixed, or when one company gets a contract whereas another one
loses it, etc.  A reasonable approach, which has a long tradition in
the economic literature
\cite{Gibrat-1931,Kalecki-1945,Champernowne-1953}, is to treat
individual income $r$ as a stochastic process and study its
probability distribution.  In general, one can study a Markov process
generated by a matrix of transitions from one income to another.  In
the case where income $r$ changes by a small amount $\Delta r$ over a
time period $\Delta t$, the Markov process can be treated as income
diffusion.  Then one can apply the general Fokker-Planck equation
\cite{Kinetics} to describe evolution in time $t$ of the income
distribution function $P(r,t)$ \cite{Yakovenko-2005a}
\begin{equation}
   \frac{\partial P}{\partial t}=\frac{\partial}{\partial r} \left[AP
   + \frac{\partial(BP)}{\partial r}\right], \; A=-{\langle\Delta
   r\rangle \over \Delta t}, \; B={\langle(\Delta r)^2\rangle \over
   2\Delta t}.
\label{diffusion}
\end{equation}
The coefficients $A$ and $B$ in Eq.\ (\ref{diffusion}) are determined
by the first and second moments of income changes per unit time.  The
stationary solution $\partial_tP=0$ of Eq.\ (\ref{diffusion}) obeys
the following equation with the general solution
\begin{equation}
  \frac{\partial(BP)}{\partial r}=-AP,\quad
  P(r)=\frac{c}{B(r)}\exp\left(-\int^r\frac{A(r')}{B(r')}dr'\right).
\label{stationary}
\end{equation}

For the lower part of the distribution, it is reasonable to assume
that $\Delta r$ is independent of $r$, i.e.,\ the changes of income are
independent of income itself.  This process is called the additive
diffusion \cite{Yakovenko-2005a}.  In this case, the coefficients in
Eq.\ (\ref{diffusion}) are constants $A_0$ and $B_0$.  Then Eq.\
(\ref{stationary}) gives the exponential distribution
$P(r)\propto\exp(-r/T_r)$ with the effective income temperature
$T_r=B_0/A_0$.  [Notice that a meaningful stationary solution
(\ref{stationary}) requires that $A>0$, i.e.,\ $\langle\Delta
r\rangle<0$.]  The coincidence of this result with the Boltzmann-Gibbs
exponential law (\ref{Gibbs}) and (\ref{money}) is not accidental.
Indeed, instead of considering pairwise interaction between particles,
one can derive Eq.\ (\ref{Gibbs}) by considering energy transfers
between a particle and a big reservoir, as long as the transfer
process is ``additive'' and does not involve a Maxwell-demon-like
discrimination.  Stochastic income fluctuations are described by a
similar process.  So, although money and income are different
concepts, they may have similar distributions, because they are
governed by similar mathematical principles.  It was shown explicitly
in Refs.\ \cite{Yakovenko-2000,Slanina-2004,Toscani-2005} that the
models of pairwise money transfer can be described in a certain limit
by the Fokker-Planck equation.

On the other hand, for the upper tail of the income distribution, it
is reasonable to expect that $\Delta r\propto r$, i.e.,\ income
changes are proportional to income itself.  This is known as the
proportionality principle of Gibrat \cite{Gibrat-1931}, and the
process is called the multiplicative diffusion \cite{Yakovenko-2005a}.
In this case, $A=ar$ and $B=br^2$, and Eq.\ (\ref{stationary}) gives
the power-law distribution $P(r)\propto1/r^{\alpha+1}$ with
$\alpha=1+a/b$.

Generally, lower-class income comes from wages and salaries, where the
additive process is appropriate, whereas upper-class income comes from
bonuses, investments, and capital gains, calculated in percentages,
where the multiplicative process applies \cite{Milakovic-2005}.
However, the additive and multiplicative processes may coexist.  An
employee may receive a cost-of-living raise calculated in percentages
(the multiplicative process) and a merit raise calculated in dollars
(the additive process).  In this case, we have $A=A_0+ar$ and
$B=B_0+br^2=b(r_0^2+r^2)$, where $r_0^2=B_0/b$.  Substituting these
expressions into Eq.\ (\ref{stationary}), we find
\begin{equation}
  P(r)=c\,\frac{e^{-(r_0/T_r)\arctan(r/r_0)}}
  {[1+(r/r_0)^2]^{1+a/2b}}.
\label{arctan}
\end{equation}
The distribution (\ref{arctan}) interpolates between the exponential
law for low $r$ and the power law for high $r$, because either the
additive or the multiplicative process dominates in the corresponding
limit.  The crossover between the two regimes takes place at $r\sim
r_0$, where the additive and multiplicative contributions to $B$ are
equal.  The distribution (\ref{arctan}) has three parameters: the
``income temperature'' $T_r=A_0/B_0$, the Pareto exponent
$\alpha=1+a/b$, and the crossover income $r_0$.  It is a minimal model
that captures the salient features of the empirical income
distribution shown in Fig.\ \ref{Fig:income1997}.  A mathematically
similar, but more economically oriented model was proposed in Refs.\
\cite{Souma-2005,Nirei-2007}, where labor income and assets
accumulation are described by the additive and a multiplicative
processes correspondingly.  A general stochastic process with additive
and multiplicative noise was studied numerically in Ref.\
\cite{Takayasu-1997}, but the stationary distribution was not derived
analytically.  A similar process with discrete time increments was
studied by Kesten \cite{Kesten-1973}.  Recently, a formula similar to
Eq.\ (\ref{arctan}) was obtained in Ref.\ \cite{Marsili}.

To verify the multiplicative and additive hypotheses empirically, it
is necessary to have data on income mobility, i.e.,\ the income
changes $\Delta r$ of the same people from one year to another.  The
distribution of income changes $P(\Delta r|r)$ conditional on income
$r$ is generally not available publicly, although it can be
reconstructed by researchers at the tax agencies.  Nevertheless, the
multiplicative hypothesis for the upper class was quantitatively
verified in Refs.\ \cite{Aoki-2003a,Aoki-2003b} for Japan, where tax
identification data are published for the top taxpayers.

The power-law distribution is meaningful only when it is limited to
high enough incomes $r>r_0$.  If all incomes $r$ from 0 to $\infty$
follow a purely multiplicative process, then one can change to a
logarithmic variable $x=\ln(r/r_*)$ in Eq.\ (\ref{diffusion}) and show
that it gives a Gaussian time-dependent distribution
$P_t(x)\propto\exp(-x^2/2\sigma^2t)$ for $x$, i.e.,\ the log-normal
distribution for $r$, also known as the Gibrat distribution
\cite{Gibrat-1931}.  However, the width of this distribution increases
linearly in time, so the distribution is not stationary.  This was
pointed out by Kalecki \cite{Kalecki-1945} a long time ago, but the
log-normal distribution is still widely used for fitting income
distribution, despite this fundamental logical flaw in its
justification.  In the classic paper \cite{Champernowne-1953},
Champernowne showed that a multiplicative process gives a stationary
power-law distribution when a boundary condition is imposed at
$r_0\neq0$.  Later, this result was rediscovered by econophysicists
\cite{Solomon-1996,Sornette-1997}.  In our Eq.\ (\ref{arctan}), the
exponential distribution of the lower class effectively provides such
a boundary condition for the power law of the upper class.  Notice
also that Eq.\ (\ref{arctan}) reduces to Eq.\ (\ref{P-Bouchaud}) in
the limit $r_0\to0$, which corresponds to purely multiplicative noise
$B=br^2$.

There are alternative approaches to income distribution in economic
literature.  One of them, proposed by Lydall \cite{Lydall-1959},
involves social hierarchy.  Groups of people have leaders, which have
leaders of a higher order, and so on.  The number of people decreases
geometrically (exponentially) with the increase of the hierarchical
level. If individual income increases by a certain factor (i.e.,\
multiplicatively) when moving to the next hierarchical level, then
income distribution follows a power law \cite{Lydall-1959}.  However,
the original argument of Lydall can be easily modified to produce the
exponential distribution.  If individual income increases by a certain
amount, i.e.,\ income increases linearly with the hierarchical level,
then income distribution is exponential.  The latter process seems to
be more realistic for moderate annual incomes below 100~k\$.  A
similar scenario is the Bernoulli trials \cite{Feller}, where
individuals have a constant probability of increasing their income by
a fixed amount.  We see that the deterministic hierarchical models and
the stochastic models of additive and multiplicative income mobility
represent essentially the same ideas.

%%%%%%%%%%%%%%%%%%%%%%%%%%%%%%%%%%%%%%%%%%%%%%%%%%%%%%%%%%%%%%%
\section{Other Applications of Statistical Physics}
\label{Sec:other}
%%%%%%%%%%%%%%%%%%%%%%%%%%%%%%%%%%%%%%%%%%%%%%%%%%%%%%%%%%%%%%%

Statistical physics was applied to a number of other subjects in
economics.  Because of lack of space, only two such topics are briefly
discussed in this section.

%%%%%%%%%%%%%%%%%%%%%%%%%%%%%%%%%%%%%%%%%%%%%%%%%%%%%%%%%%%%%%%
\subsection{Economic temperatures in different countries}
\label{Sec:temperatures}
%%%%%%%%%%%%%%%%%%%%%%%%%%%%%%%%%%%%%%%%%%%%%%%%%%%%%%%%%%%%%%%

As discussed in Secs.\ \ref{Sec:w-empirical} and \ref{Sec:r-data}, the
distributions of money, wealth, and income are often described by
exponential functions for the majority of the population.  These
exponential distributions are characterized by the parameters $T_m$,
$T_w$, and $T_r$, which are mathematically analogous to temperature in
the Boltzmann-Gibbs distribution (\ref{Gibbs}).  The values of these
parameters, extracted from the fits of the empirical data, are
generally different for different countries, i.e.,\ different
countries have different economic ``temperatures''.  For example,
Ref.\ \cite{Yakovenko-2001b} found that the US income temperature was
1.9 times higher than the UK income temperature in 1998 (using the
exchange rate of dollars to pounds at that time).  Also, there was
$\pm25\%$ variation between income temperatures of different states
within the USA \cite{Yakovenko-2001b}.

In physics, a difference of temperatures allows one to set up a
thermal machine.  In was argued in Ref.\ \cite{Yakovenko-2000} that
the difference of money or income temperatures between different
countries allows one to extract profit in international trade.
Indeed, as discussed at the end of Sec.\ \ref{Sec:w-empirical}, the
prices of goods should be commensurate with money or income
temperature, because otherwise people cannot afford to buy those
goods.  So, an international trading company can buy goods at a low
price $T_1$ in a ``low-temperature'' country and sell them at a high
price $T_2$ in a ``high-temperature'' country.  The difference of
prices $T_2-T_1$ would be the profit of the trading company.  In this
process, money (the analog of energy) flows from the
``high-temperature'' to the ``low-temperature'' country, in agreement
with the second law of thermodynamics, whereas products flow in the
opposite direction.  This process very much resembles what is going on
in global economy now.  In this framework, the perpetual trade deficit
of the USA is the consequence of the second law of thermodynamics and
the difference of temperatures between the USA and the
``low-temperature'' countries, such as China.  Similar ideas were
developed in more detail in Refs.\ \cite{Mimkes-2005b,Mimkes-2006a},
including a formal Carnot cycle for international trade.

The statistical physics approach demonstrates that profit originates
from statistical nonequilibrium (the difference of temperatures),
which exists in the global economy.  However, it does not answer the
question what is the origin of this difference.  By analogy with
physics, one would expect that the money flow should reduce the
temperature difference and, eventually, lead to equilibrization of
temperatures.  In physics, this situation is known as the ``thermal
death of the universe''.  In a completely equilibrated global economy,
it would be impossible to make profit by exploiting differences of
economic temperatures between different countries.  Although
globalization of modern economy does show a tendency toward
equilibrization of living standards in different countries, this
process is far from straightforward, and there are many examples
contrary to equilibrization.  This interesting and timely subject
certainly requires further study.

%%%%%%%%%%%%%%%%%%%%%%%%%%%%%%%%%%%%%%%%%%%%%%%%%%%%%%%%%%%%%%%
\subsection{Society as a binary alloy}
\label{Sec:alloy}
%%%%%%%%%%%%%%%%%%%%%%%%%%%%%%%%%%%%%%%%%%%%%%%%%%%%%%%%%%%%%%%

In 1971, Thomas Schelling proposed the now-famous mathematical model
of segregation \cite{Schelling-1971}.  He considered a lattice, where
the sites can be occupied by agents of two types, e.g.,\ blacks and
whites in the problem of racial segregation.  He showed that, if the
agents have some probabilistic preference for the neighbors of the
same type, the system spontaneously segregates into black and white
neighborhoods.  This mathematical model is similar to the so-called
Ising model, which is a popular model for studying phase transitions
in physics.  In this model, each lattice site is occupied by a
magnetic atom, whose magnetic moment has only two possible
orientations, up or down.  The interaction energy between two
neighboring atoms depends on whether their magnetic moments point in
the same or in the opposite directions.  In physics language, the
segregation found by Schelling represents a phase transition in this
system.

Another similar model is the binary alloy, a mixture of two elements
which attract or repel each other.  It was noticed in Ref.\
\cite{Mimkes-1995} that the behavior of actual binary alloys is
strikingly similar to social segregation.  In the following papers
\cite{Mimkes-2000,Mimkes-2006b}, this mathematical analogy was
developed further and compared with social data.  Interesting
concepts, such as the coexistence curve between two phases and the
solubility limit, were discussed in this work.  The latter concept
means that a small amount of one substance dissolves into another up
to some limit, but phase separation (segregation) develops for higher
concentrations.  Recently, similar ideas were rediscovered in Refs.\
\cite{Roehner-2007,Stauffer-2007,Castellano-2008}.  The vast
experience of physicists in dealing with phase transitions and alloys
may be helpful for practical applications of such models
\cite{Bar-Yam-2007}.

%%%%%%%%%%%%%%%%%%%%%%%%%%%%%%%%%%%%%%%%%%%%%%%%%%%%%%%%%%%%%%%
\section{Future Directions, Criticism, and Conclusions}
%%%%%%%%%%%%%%%%%%%%%%%%%%%%%%%%%%%%%%%%%%%%%%%%%%%%%%%%%%%%%%%

The statistical models described in this review are quite simple.  It
is commonly accepted in physics that theoretical models are not
intended to be photographic copies of reality, but rather be
caricatures, capturing the most essential features of a phenomenon
with a minimal number of details.  With only few rules and parameters,
the models discussed in Secs.\ \ref{Sec:money}, \ref{Sec:wealth}, and
\ref{Sec:income} reproduce spontaneous development of stable
inequality, which is present in virtually all societies.  It is
amazing that the calculated Gini coefficients, $G=1/2$ for individuals
and $G=3/8$ for families, are actually very close to the US income
data, as shown in Fig.\ \ref{Fig:Lorenz} and \ref{Fig:Lorenz-2}.
These simple models establish a baseline and a reference point for
development of more sophisticated and more realistic models.  Some of
these future directions are outlined below.

%%%%%%%%%%%%%%%%%%%%%%%%%%%%%%%%%%%%%%%%%%%%%%%%%%%%%%%%%%%%%%%
\subsection{Future directions}
\label{Sec:future}
%%%%%%%%%%%%%%%%%%%%%%%%%%%%%%%%%%%%%%%%%%%%%%%%%%%%%%%%%%%%%%%

\paragraph{Agents with a finite lifespan.}

The models discussed in this review consider immortal agents who live
forever, like atoms.  However, humans have a finite lifespan.  They
enter the economy as young people and exit at an old age.  Evolution
of income and wealth as functions of age is studied in economics using
the so-called overlapping-generations model.  The absence of the age
variable was one of the criticisms of econophysics by the economist
Paul Anglin \cite{Anglin-2005}.  However, the drawback of the standard
overlapping-generations model is that there is no variation of income
and wealth between agents of the same age, because it is a
representative-agent model.  It would be best to combine stochastic
models with the age variable.  Also, to take into account inflation of
average income, Eq.\ (\ref{diffusion}) should be rewritten for
relative income, in the spirit of Eq.\ (\ref{FP-Bouchaud}).  These
modifications would allow to study the effects of demographic waves,
such as baby boomers, on the distributions of income and wealth.

\paragraph{Agent-based simulations of the two-class society.}

The empirical data presented in Sec.\ \ref{Sec:r-data} show quite
convincingly that the US population consists of two very distinct
classes characterized by different distribution functions.  However,
the theoretical models discussed in Secs.\ \ref{Sec:money} and
\ref{Sec:wealth} do not produce two classes, although they do produce
broad distributions.  Generally, not much attention has been payed in
the agent-based literature to simulation of two classes.  One
important exception is Ref.\ \cite{Wright-2005}, in which spontaneous
development of employers and employees classes from initially equal
agents was simulated \cite{Wright-2007}.  More work in this direction
would be certainly desirable.

\paragraph{Access to detailed empirical data.}

A great amount of statistical information is publicly available on the
Internet, but not for all types of data.  As discussed in Sec.\
\ref{Sec:w-empirical}, it would be very interesting to obtain data on
the distribution of balances on bank accounts, which would give
information about the distribution of money (as opposed to wealth).
As discussed in Sec.\ \ref{Sec:r-theory}, it would be useful to obtain
detailed data on income mobility, to verify the additive and
multiplicative hypotheses for income dynamics.  Income distribution is
often reported as a mix of data on individual income and family
income, when the counting unit is a tax return (joint or single) or a
household.  To have a meaningful comparison with theoretical models,
it is desirable to obtain clean data where the counting unit is an
individual.  Direct collaboration with statistical agencies would be
very useful.

\paragraph{Economies in transition.}

Inequality in developed capitalist countries is generally quite
stable.  The situation is very different for the former socialist
countries making a transition to a market economy.  According to the
World Bank data \cite{Yakovenko-2003}, the average Gini coefficient
for family income in eastern Europe and the former Soviet Union jumped
from 25\% in 1988 to 47\% in 1993.  The Gini coefficient in the
socialist countries before the transition was well below the
equilibrium value of 37.5\% for market economies.  However, the fast
collapse of socialism left these countries out of market equilibrium
and generated a much higher inequality.  One may expect that, with
time, their inequality will decrease to the equilibrium value of
37.5\%.  It would be very interesting to trace how fast this
relaxation takes place.  Such a study would also verify whether the
equilibrium level of inequality is universal for all market economies.

\paragraph{Relation to physical energy.}

The analogy between energy and money discussed in Sec.\
\ref{Sec:conservation} is a formal mathematical analogy.  However,
actual physical energy with low entropy (typically in the form of
fossil fuel) also plays a very important role in the modern economy,
being the basis of current human technology.  In view of the looming
energy and climate crisis, it is imperative to find realistic ways for
making a transition from the current ``disposable'' economy based on
``cheap'' and ``unlimited'' energy and natural resources to a
sustainable one.  Heterogeneity of human society is one of the
important factors affecting such a transition.  Econophysics, at the
intersection of energy, entropy, economy, and statistical physics, may
play a useful role in this quest \cite{Defilla-2007}.

%%%%%%%%%%%%%%%%%%%%%%%%%%%%%%%%%%%%%%%%%%%%%%%%%%%%%%%%%%%%%%%
\subsection{Criticism from economists}
\label{Sec:criticism}
%%%%%%%%%%%%%%%%%%%%%%%%%%%%%%%%%%%%%%%%%%%%%%%%%%%%%%%%%%%%%%%
\setcounter{paragraph}{0}

As econophysics is gaining popularity, some criticism has appeared
from economists \cite{Anglin-2005}, including those who are closely
involved with the econophysics movement
\cite{Lux-2005,Ormerod-2006,Lux-2008}.  This reflects a long-standing
tradition in economic and social sciences of writing critiques on
different schools of thought.  Much of the criticism is useful and
constructive and is already being accommodated in the econophysics
work.  However, some criticism results from misunderstanding or
miscommunication between the two fields and some from significant
differences in scientific philosophy.  Several insightful responses to
the criticism have been published
\cite{McCauley-2006,response-2006,Rosser-2008b}, see also
\cite{Stauffer-history,Rosser-2006a}.  In this section, we briefly
address the issues that are directly related to the material discussed
in this review.

\paragraph{Awareness of previous economic literature.}

One complaint of Refs.\
\cite{Anglin-2005,Lux-2005,Ormerod-2006,Lux-2008} is that physicists
are not well aware of the previous economic literature and either
rediscover known results or ignore well-established approaches.  To
address this issue, it is useful to keep in mind that science itself
is a complex system, and scientific progress is an evolutionary
process with natural selection.  The sea of scientific literature is
enormous, and nobody knows it all.  Recurrent rediscovery of
regularities in the natural and social world only confirms their
validity.  Independent rediscovery usually brings a different
perspective, broader applicability range, higher accuracy, and better
mathematical treatment, so there is progress even when some overlap
with previous results exists.  Physicists are grateful to economists
for bringing relevant and specific references to their attention.
Since the beginning of modern econophysics, many old references have
been uncovered and are now routinely cited.

However, not all old references are relevant to the new development.
For example, Ref.\ \cite{Ormerod-2006} complained that the
econophysics literature on income distribution ignores the so-called
Kuznets hypothesis \cite{Kuznets-1955}.  The Kuznets hypothesis
postulates that income inequality first rises during an industrial
revolution and then decreases, producing an inverted-U-shaped curve.
Ref.\ \cite{Ormerod-2006} admitted that, to date, the large amount of
literature on the Kuznets hypothesis is inconclusive.  Ref.\
\cite{Ormerod-2006} mentioned that this hypothesis applies to the
period from colonial times to 1970s; however, the empirical data for
this period are sparse and not very reliable.  The econophysics
literature deals with the reliable volumes of data for the second half
of the 20th century, collected with the introduction of computers.  It
is not clear what is the definition of industrial revolution and when
exactly it starts and ends.  The chain of technological progress seems
to be continuous (steam engine, internal combustion engine, cars,
plastics, computers, Internet), so it is not clear where the purported
U-curve is supposed to be placed in time.  Thus, the Kuznets
hypothesis appears to be, in principle, unverifiable and
unfalsifiable.  The original paper by Kuznets \cite{Kuznets-1955}
actually does not contain any curves, but it has one table filled with
made-up, imaginary data!  Kuznets admits that he has ``neither the
necessary data nor a reasonably complete theoretical model'' \cite[p
12]{Kuznets-1955}.  So, this paper is understandably ignored by the
econophysics literature.  In fact, the data analysis for 1947--1984
shows amazing stability of income distribution \cite{Levy-1987},
consistent with Fig.\ \ref{Fig:Lorenz-2}.  The increase of inequality
in the 1990s resulted from growth of the upper tail relative to the
lower class, but the relative inequality within the lower class
remains very stable, as shown in Fig.\ \ref{Fig:income}.

\paragraph{Reliance on visual data analysis.}

Another complaint of Ref.\ \cite{Ormerod-2006} is that econophysicists
favor graphic analysis of data over the formal and ``rigorous''
testing prescribed by mathematical statistics, as favored by
economists.  This complaint goes against the trend of all sciences to
use increasingly sophisticated data visualization for uncovering
regularities in complex system.  The thick IRS publication 1304
\cite{Pub1304} is filled with data tables, but has virtually no
graphs.  Despite the abundance of data, it gives a reader no idea
about income distribution, whereas plotting the data immediately gives
insight.  However, intelligent plotting is the art with many tools,
which not many researchers have mastered.  The author completely
agrees with Ref.\ \cite{Ormerod-2006} that too many papers mindlessly
plot any kind of data on a log-log scale, pick a finite interval,
where any smooth curved line can be approximated by a straight line,
and claim that there is a power law.  In many cases, replotting the
same data on a log-linear scale converts a curved line into a straight
line, which means that the law is actually exponential.

Good visualization is extremely helpful in identifying trends in
complex data, which can then be fitted to a mathematical function.
However, for a complex system, such a fit should not be expected with
infinite precision.  The fundamental laws of physics, such as Newton's
law of gravity or Maxwell's equations, are valid with enormous
precision.  However, the laws in condensed matter physics, uncovered
by experimentalists with a combination of visual analysis and fitting,
usually have much lower precision, at best 10\% or so.  Most of these
laws would fail the formal criteria of mathematical statistics.
Nevertheless these approximate laws are enormously useful in practice,
and the everyday devices, engineered on the basis of these laws, work
very well for all of us.

Because of the finite accuracy, different functions may produce
equally good fits.  Discrimination between the exponential, Gamma, and
log-normal functions may not be always possible \cite{Yakovenko-2006}.
However, the exponential function has fewer fitting parameters, so it
is preferable on the basis of simplicity.  The other two functions can
simply mimic the exponential function with a particular choice of the
additional parameters \cite{Yakovenko-2006}.  Unfortunately, many
papers in mathematical statistics introduce too many fitting
parameters into complicated functions, such as the generalized beta
distribution mentioned in Ref.\ \cite{Ormerod-2006}.  Such
overparametrization is more misleading than insightful for data
fitting.

\paragraph{Quest for universality.}

Ref.\ \cite{Ormerod-2006} criticized physicists for trying to find
universality in economic data.  It also seems to equate the concepts
of power law, scaling, and universality.  These are three different,
albeit overlapping, concepts.  Power laws usually apply only to a
small fraction of data at the high ends of various distributions.
Moreover, the exponents of these power laws are usually nonuniversal
and vary from case to case.  Scaling means that the shape of a
function remains the same when its scale changes.  However, the
scaling function does not have to be a power-law function.  A good
example of scaling is shown in Fig.\ \ref{Fig:income}, where income
distributions for the lower class collapse on the same exponential
line for about 20 years of data.  We observe amazing universality of
income distribution, unrelated to a power law.  In a general sense,
the diffusion equation is universal, because it describes a wide range
of systems, from dissolution of sugar in water to a random walk in the
stock market.

Universalities are not easy to uncover, but they form the backbone of
regularities in the world around us.  This is why physicists are so
much interested in them.  Universalities establish the first-order
effect, and deviations represent the second-order effect.  Different
countries may have somewhat different distributions, and economists
often tend to focus on these differences.  However, this focus on
details misses the big picture that, in the first approximation, the
distributions are quite similar and universal.

\paragraph{Theoretical modeling of money, wealth, and income.}

Refs.\ \cite{Anglin-2005,Ormerod-2006,Lux-2008} pointed out that many
econophysics papers confuse or misuse the terms for money, wealth, and
income.  It is true that terminology is sloppy in many papers and
should be refined.  However, the terms in Refs.\
\cite{Yakovenko-2000,Chakraborti-2000} are quite precise, and this
review clearly distinguishes between these concepts in Secs.\
\ref{Sec:money}, \ref{Sec:wealth}, and \ref{Sec:income}.

One contentious issue is about conservation of money.  Ref.\
\cite{Ormerod-2006} agrees that ``transactions are a key economic
process, and they are necessarily conservative'', i.e.,\ money is
indeed conserved in transactions between agents.  However, Refs.\
\cite{Anglin-2005,Ormerod-2006,Lux-2008} complain that the models of
conservative exchange do not consider production of goods, which is
the core economic process and the source of economic growth.  Material
production is indeed the ultimate goal of the economy, but it does not
violate conservation of money by itself.  One can grow coffee beans,
but nobody can grow money on a money tree.  Money is an artificial
economic device that is designed to be conserved.  As explained in
Sec.\ \ref{Sec:money}, the money transfer models implicitly assume
that money in transactions is voluntarily payed for goods and services
generated by production for the mutual benefit of the parties.  In
principle, one can introduce a billion of variables to keep track of
every coffee bean and other product of the economy.  What difference
would it make for the distribution of \emph{money}?  Despite claims in
Refs.\ \cite{Anglin-2005,Ormerod-2006}, there is no contradiction
between models of conservative exchange and the classic work of Adam
Smith and David Ricardo.  The difference is only in the focus: We keep
track of money, whereas they keep track of coffee beans, from
production to consumption.  These approaches address different
questions, but do not contradict each other.  Because money constantly
circulates in the system as payment for production and consumption,
the resulting statistical distribution of money may very well not
depend on what is exactly produced and in what quantities.

In principle, the models with random transfers of money should be
considered as a reference point for developing more sophisticated
models.  Despite the totally random rules and ``zero intelligence'' of
the agents, these models develop well-characterized, stable and
stationary distributions of money.  One can modify the rules to make
the agents more intelligent and realistic and see how much the
resulting distribution changes relative to the reference one.  Such an
attempt was made in Ref.\ \cite{Lux-2005} by modifying the model of
Ref.\ \cite{Silver-2002} with various more realistic economic
ingredients.  However, despite the modifications, the resulting
distributions were essentially the same as in the original model.
This example illustrates the typical robustness and universality of
statistical models: Modifying details of microscopic rules does not
necessarily change the statistical outcome.

Another misconception, elaborated in Ref.\ \cite{Lux-2005,Lux-2008},
is that the money transfer models discussed in Sec.\ \ref{Sec:money}
imply that money is transferred by fraud, theft, and violence, rather
than voluntarily.  One should keep in mind that the catchy labels
``theft-and-fraud'', ``marriage-and-divorce'', and ``yard-sale'' were
given to the money transfer models by the journalist Brian Hayes in a
popular article \cite{Hayes-2002}.  Econophysicists who originally
introduced and studied these models do not subscribe to this
terminology, although the early work of Angle \cite{Angle-1986} did
mention violence as one source of redistribution.  In the opinion of
the author, it is indeed difficult to justify the proportionality rule
(\ref{proportional}), which implies that agents with high balances pay
proportionally greater amounts in transactions than agents with low
balances.  However, the additive model of Ref.\ \cite{Yakovenko-2000},
where money transfers $\Delta m$ are independent of money balances
$m_i$ of the agents, does not have this problem.  As explained in
Sec.\ \ref{Sec:BGmoney}, this model simply means that all agents pay
the same prices for the same product, although prices may be different
for different products.  So, this model is consistent with voluntary
transactions in a free market.

Ref.\ \cite{McCauley-2006} argued that conservation of money is
violated by credit.  As explained in Sec.\ \ref{Sec:debt}, credit does
not violate conservation law, but creates positive and negative money
without changing net worth.  Negative money (debt) is as real as
positive money.  Ref.\ \cite{McCauley-2006} claimed that money can be
easily created with the tap of a computer key via credit.  Then why
would an employer not tap the key and double salaries, or a funding
agency double research grants?  Because budget constraints are real.
Credit may provide a temporary relief, but sooner or later it has to
be paid back.  Allowing debt may produce a double-exponential
distribution as shown in Fig.\ \ref{Fig:reserve}, but it does not
change the distribution fundamentally.

As discussed in Sec.\ \ref{Sec:conservation}, a central bank or a
central government can inject new money into the economy.  As
discussed in Sec.\ \ref{Sec:wealth}, wealth is generally not
conserved.  As discussed in Sec.\ \ref{Sec:income}, income is
different from money and is described by a different model
(\ref{diffusion}).  However, the empirical distribution of income
shown in Fig.\ \ref{Fig:income1997} is qualitatively similar to the
distribution of wealth shown in Fig.\ \ref{Fig:UKwealth}, and we do
not have data on money distribution.

%%%%%%%%%%%%%%%%%%%%%%%%%%%%%%%%%%%%%%%%%%%%%%%%%%%%%%%%%%%%%%%
\subsection{Conclusions}
\label{Sec:conclusions}
%%%%%%%%%%%%%%%%%%%%%%%%%%%%%%%%%%%%%%%%%%%%%%%%%%%%%%%%%%%%%%%

The ``invasion'' of physicists into economics and finance at the turn
of the millennium is a fascinating phenomenon.  The physicist Joseph
McCauley proclaims that ``Econophysics will displace economics in both
the universities and boardrooms, simply because what is taught in
economics classes doesn't work'' \cite{Ball-2006}.  Although there is
some truth in his arguments \cite{McCauley-2006}, one may consider a
less radical scenario.  Econophysics may become a branch of economics,
in the same way as games theory, psychological economics, and now
agent-based modeling became branches of economics.  These branches
have their own interests, methods, philosophy, and journals.  The main
contribution from the infusion of new ideas from a different field is
not in answering old questions, but in raising new questions.  Much of
the misunderstanding between economists and physicists happens not
because they are getting different answers, but because they are
answering different questions.

The subject of income and wealth distributions and social inequality
was very popular at the turn of another century and is associated with
the names of Pareto, Lorenz, Gini, Gibrat, and Champernowne, among
others.  Following the work by Pareto, attention of researchers was
primarily focused on the power laws.  However, when physicists took a
fresh, unbiased look at the empirical data, they found a different,
exponential law for the lower part of the distribution.  The
motivation for looking at the exponential law, of course, came from
the Boltzmann-Gibbs distribution in physics.  Further studies provided
a more detailed picture of the two-class distribution in a society.
Although social classes have been known in political economy since
Karl Marx, realization that they are described by simple mathematical
distributions is quite new.  Demonstration of the ubiquitous nature of
the exponential distribution for money, wealth, and income is one of
the new contributions produced by econophysics.

%%%%%%%%%%%%%%%%%%%%%%%%%%%%%%%%%%%%%%%%%%%%%%%%%%%%%%%%%%%%%%%

%%%%%%%%%%%%%%%%%%%%%%%%%%%%%%%%%%%%%%%%%%%%%%%%%%%%%%%%%%%%%%%%
\section*{Books and Reviews}
%%%%%%%%%%%%%%%%%%%%%%%%%%%%%%%%%%%%%%%%%%%%%%%%%%%%%%%%%%%%%%%%

\begin{itemize}

\item McCauley J (2004) Dynamics of Markets: Econophysics and
Finance. Cambridge University Press, Cambridge

\item Farmer JD, Shubik M, Smith E (2005) Is economics the next
physical science?  Physics Today 58(9):37-42

\item Samanidou E, Zschischang E, Stauffer D, Lux T (2007) Agent-based
models of financial markets.  Reports on Progress in Physics
70:409--450

\item Web resource: Econophysics Forum
\url{http://www.unifr.ch/econophysics/}

\end{itemize}

%%%%%%%%%%%%%%%%%%%%%%%%%%%%%%%%%%%%%%%%%%%%%%%%%%%%%%%%%%%%%%%%
%%%%%%%%%%%%%%%%%%%%%%%%%%%%%%%%%%%%%%%%%%%%%%%%%%%%%%%%%%%%%%%%
\end{document}